\documentclass[aos,preprint]{imsart}
\RequirePackage[OT1]{fontenc}
\RequirePackage{amsthm,amsmath}
\RequirePackage{natbib}
\RequirePackage[colorlinks,citecolor=blue,urlcolor=blue]{hyperref}
\arxiv{arXiv:0000.0000}
\startlocaldefs
\numberwithin{equation}{section}
\theoremstyle{plain}
\endlocaldefs
\usepackage{amsmath, mathrsfs, amsthm}
\usepackage{graphicx}
\usepackage{verbatim}
\usepackage{float}

\newtheorem{lemma}{Lemma}
\newtheorem{cor}{Corollary}

\newtheorem{definition}{Definition}

\newcommand{\bed}{\begin{definition}}
\newcommand{\eed}{\end{definition}}
\newcommand{\beq}{\begin{equation}}
\newcommand{\eeq}{\end{equation}}

\newtheorem{theorem}{Theorem}

\newcommand{\eps}{\epsilon}

\newcommand{\bitem}{\begin{itemize}}
\newcommand{\eitem}{\end{itemize}}

\newcommand{\goto}{\rightarrow}

\newcommand{\beqn}{\begin{equation}}
\newcommand{\eeqn}{\end{equation}}
\newcommand{\balign}{\begin{align}}
\newcommand{\ealign}{\end{align}}

\newcommand{\cG}{{\cal G}}

\newcommand{\tv}{\tilde{v}}

\newcommand{\cN}{{\cal N}}

\newcommand{\ty}{\tilde{y}}

\newcommand{\call}{{\cal I}}

\def\limsup{\mathop{\overline{\rm lim}}}

\newcommand{\bel}{\begin{eqnarray}\label}
\newcommand{\eel}{\end{eqnarray}}
\newcommand{\bes}{\begin{eqnarray*}}
\newcommand{\ees}{\end{eqnarray*}}

\def\tX{{\tilde X}}
\def\tE{{\tilde E}}

\def\cA{{\cal A}}

\newcommand{\tcall}{\tilde{\cal I}}
\newcommand{\cJ}{{\cal J}}

\usepackage{amssymb}
\usepackage{color}
\usepackage{multicol}
\usepackage{subfigure}
\usepackage{multirow}
\usepackage{enumitem}
\newcommand{\rom}[1]{%
  \textup{\uppercase\expandafter{\romannumeral#1}}%
}

\begin{document}

\begin{frontmatter}
\title{Covariate Assisted Variable Ranking}
\runtitle{Covariate Assisted Variable Ranking}

\begin{aug}
\author{\fnms{Zheng Tracy} \snm{Ke}\thanksref{t1}\ead[label=e1]{zke@galton.uchicago.edu}}
\and
\author{\fnms{Fan} \snm{Yang}\ead[label=e2]{fyang1@galton.uchicago.edu}}
\thankstext{t1}{Supported in part by NSF Grant DMS-1712958.}

\runauthor{Z. Ke and F. Yang}

\affiliation{University of Chicago}

\address{
Z. Ke and F. Yang\\
Department of Statistics\\
University of Chicago\\
Chicago, Illinois, IL 60637\\
\printead{e1}\\
\phantom{E-mail:\ }\printead*{e2}}

\end{aug}

\begin{abstract}
Consider a linear model $y = X \beta + z$,  $z \sim N(0,  \sigma^2 I_n)$.  
The Gram matrix $\Theta = \frac{1}{n} X'X$ is non-sparse, but it 
is approximately the sum of two components,  a low-rank matrix and a sparse matrix, where 
neither component is known to us. 
We are interested in the Rare/Weak signal setting where all but  
a small fraction of the entries of $\beta$ are nonzero, and the nonzero entries
are relatively small individually.   The goal is to rank the variables in a way so as to maximize the area under the ROC curve.  

We propose {\it Factor-adjusted Covariate Assisted Ranking} (FA-CAR) as a two-step approach 
to variable ranking. In the FA-step, we use PCA to reduce the linear model to a new one 
where the Gram matrix is approximately sparse.  In the 
CAR-step, we rank variables by exploiting the local covariate structures.  

FA-CAR is easy to use and computationally fast, and it is effective in resolving signal cancellation, a challenge we face in regression models. 
FA-CAR is related to the recent idea of Covariate Assisted Screening and Estimation (CASE),  but 
two methods are for different goals and are thus very different. 

We compare the ROC curve of FA-CAR with some other ranking ideas on numerical experiments, and show that FA-CAR has several advantages. 
Using a Rare/Weak signal model,  we derive the convergence rate of the minimum sure-screening model size of FA-CAR.  
Our theoretical analysis contains several new ingredients, especially a new perturbation bound for PCA.

\end{abstract}



\end{frontmatter}

\section{Introduction} \label{sec:intro}
Consider the linear model in the $p \gg n$ setting: 
\beq \label{linearmod}
y = X\beta + z, \qquad X=[x_1,\cdots, x_p]\in\mathbb{R}^{n,p},\qquad z\sim N(0, \sigma^2 I_n). 
\eeq
We are interested in variable ranking, a problem related to variable selection but very different. Scientific experiments are constrained by budget and manpower,  and it is often impossible to completely separate the signals from the noise.   An  
alternative is then to identify a few most promising variables for  follow-up lab experiments. This is where variable ranking comes in.

In this paper, we call a nonzero entry of $\beta$ a ``signal" and a zero entry a ``noise." 
We are interested in the Rare/Weak signal regime:  
\begin{itemize}
\item ({\it Rare}).  All but a small fraction of the entries of $\beta$ are nonzero. 
\item ({\it Weak}). Individually, the nonzero entries are relatively small.  
\end{itemize} 
We assume the Gram matrix $\Theta=(1/n)X'X$ follows an {\it approximate factor model} \citep{chamberlain1983arbitrage,FLM-poet1}:
\beq \label{mod-G}
\Theta = \sum_{k=1}^K \lambda_k v_k v_k' + G_0,   \qquad K \ll \min\{n, p\}, 
\eeq 
where $G_0$ is positive definite and sparse (in the sense that each row has relatively few large entries, 
with all other entries relatively small),   $\lambda_1,\cdots, \lambda_K$ are positive, and 
$v_1,\cdots,v_K$ are mutually orthogonal unit-norm vectors. 

Model (\ref{mod-G})  is popular in finance \citep{connor1993test}, where the low-rank part represents a few   risk factors and $G_0$ is the covariance matrix of the (weakly-correlated) idiosyncratic noise. It is also useful in microarray analysis, where the low-rank part represents 
 technical, environmental, demographic, or genetic factors
\citep{leek2007capturing}.  

Marginal Ranking (MR) is an approach that is especially popular in genomics and genetics \citep{jeffery2006comparison, liu2002comparative}.  Recall that $X = [x_1, x_2, \ldots, x_p]$. The method ranks variables according to the marginal regression coefficients $|(x_j, y)|/(x_j, x_j)$, $1 \leq j \leq p$,   where $(\cdot, \cdot)$ denotes the inner product. Marginal ranking has advantages: (a) It directly provides an explicit ranking that is reasonable in many cases; (b) It is easy-to-use and computationally fast; (c) It does not need tuning parameters; (d) It has a relatively less stringent requirement on noise distributions and provides 
reasonable results even when the noise distribution is unknown or when the features/samples are correlated.       

Penalized methods are well-known variable selection approaches \citep{lasso,SCAD,zou2006adaptive}. 
However, compared to MR, penalization approaches are different in all the above aspects:  (a) Variable selection and variable ranking are two different goals, and penalization methods do not automatically provide an explicit ranking; (b) They are  computationally slower,  especially when $(n, p)$ are large; (c) They usually require tuning; (d) They require more stringent conditions 
on the noise distribution; particularly, their behavior deteriorates when the noise distribution is misspecified \citep{wu2016performance}. 

Our goal is to improve MR so that it works well in our setting. Despite many good aspects of MR, we recognize that it faces two challenges:  
\begin{itemize} 
\item The $K$ factors in $\Theta$ may have a dominant effect, and  
the ranking by MR is only reasonable when these factors are removed. 
\item MR faces the so-called challenge of ``signal cancellation" \citep{WassermanRoeder}.
\end{itemize} 
Note that 
\[ 
E[(x_j, y)/(x_j, x_j)] = (x_j, x_j)^{-1} \sum_{k: \beta_k \neq 0} (x_j, x_k) \beta_k. 
\] 
``Signal cancellation" means that due to correlations among $x_j$'s,   signals may have a mutual canceling effects, and variable $j$ may receive a relatively low ranking even when $\beta_j$ is top-ranked among $\beta_1, \beta_2, \ldots, \beta_p$.

To overcome the challenges, we propose FA-CAR as a two-stage ranking method. 
FA-CAR contains a Factor-Adjusting (FA) step, where we use PCA 
for factor removal and reduce the linear 
model to a new one where the Gram matrix is sparse.   
In the Covariate-Assisted Ranking (CAR) step, we rank variables 
using covariate structures. We recognize that ``signal cancellation" is only 
severe when the predictors are heavily correlated, and so by exploiting the 
covariate structures, we can significantly alleviate the canceling effects.  
Our major contributions are 
\begin{itemize} 
\item {\it (A new ranking method).} FA-CAR is easy to use and computationally fast, and it is effective in resolving ``signal cancellation." The numerical comparison of ROC curves shows that our method has advantages over some existing ranking methods. 
\item {\it (Rate of convergence).} Using a Rare/Weak signal model, we derive the convergence rate of the minimum sure-screening model size of FA-CAR. The advantage of FA-CAR is validated by the theoretical results. 
\item {\it (New technical tools).}  In our analysis, we develop some new technical tools. Particularly, the FA step requires sharp perturbation analysis of PCA (Section~\ref{subsec:FA-thm}), which is new to the best of our knowledge. 
\item {\it (Extensions to GLM).}  We extend FA-CAR to a variable ranking method for generalized linear models (GLM). 
\end{itemize}

\subsection{Two illustrating examples} \label{subsec:idea} 
It is instructive to use  two simple examples to illustrate why FA and CAR are useful.

{\bf Example 1} ({\it One-factor design}).   Consider a case where the Gram matrix 
 \[
 \Theta= I_p + \omega_p  \xi \xi', \qquad \mbox{$\omega_p>0$ is a parameter},     
 \] 
where $\xi  = \eta/\|\eta\|$ with $\eta \sim N(0, I_n)$ and  $\sigma^2 = 0$ so there is no noise.  
We assume $\beta$ has $s$ nonzeros and each nonzero equals to $\tau$ ($\tau > 0$). 
Even in this simple setting, many methods do not perform well. Take MR  for example. As long as $
w_p \ll p$, we have $n^{-1}(x_j, x_j) \approx 1$ and    
\[
|(x_j,y)|/(x_j,x_j)  \sim |\omega_p  \cdot (\xi, \beta) \xi_j  +  \beta_j|.  
\]  
Since $|(\xi, \beta)\xi_j| = O_p(\tau \sqrt{s}/p )$, whenever $w_p \sqrt{s}/p \gg 1$, the factor has a non-negligible  effect:  the ranking depends more on  $\xi$ instead of $\beta$, and many signal variables may receive lower rankings than the noise variables.

Seemingly, the problem can be fixed if we use a factor removal step. Consider the Singular Vector Decomposition (SVD) of the design matrix $X$: 
\[
X = \sum_{k = 1}^n \lambda_k u_k v_k' \equiv \lambda_1 u_1 v_1' + \tilde{X},      \qquad \mbox{where} \;  v_1 = \xi  \;  \mbox{and}  \;   \tilde{X} = \sum_{k = 2}^n \lambda_k u_k v_k'. 
\] 
We have two ways to remove the factor $\xi$: one is to project the columns of $X$ using the projection matrix $H_{u} = I_n - u_1 u_1'$, and the other one is to project the rows of $X$ using the 
projection matrix $H_v = I_p - v_1 v_1'$. However, while both projections produce the same matrix:   
\[
\tilde{X}  = H_u X = X H_v, 
\] 
only the first one reduces Model (\ref{linearmod}) to a new linear model with the same vector $\beta$. In particular, letting $\tilde{y} = H_u y$ and $\tilde{\eps} = H_u  \eps$, we have 
\[
\tilde{y} = \tilde{X} \beta + \tilde{z}, \qquad \mbox{where} \;  \tilde{z} \sim N(0, \sigma^2 H_u)  \; \mbox{and} \; n^{-1}\tilde{X}' \tilde{X} = H_v =  I_p - \xi \xi'. 
\] 
Similarly, if we write $\tilde{X} = [\tilde{x}_1, \tilde{x}_2, \ldots, \tilde{x}_p]$ and apply MR, then $n^{-1}(\tilde{x}_j, \tilde{x}_j) \approx 1$ and 
\[
|(\tilde{y}, \tilde{x}_j)|/(\tilde{x}_j, \tilde{x}_j)  \sim | (\xi, \beta) \xi_j + \beta_j|, 
\]
where $|(\xi, \beta) \xi_j| = O_p(\tau \sqrt{s} / p)$ and has a negligible effect on the ranking. 
We therefore have a successful ranking scheme if we first remove the factor and then apply MR (in this simple setting, the sparse component $G_0$ in $\Theta$ is diagonal). 
This is the basic idea of the FA step, which can be conveniently extended to cases where we have 
more than $1$ factors.

{\bf Example 2}.  ({\it Block-wise diagonal design}). Suppose $p$ is even and $\Theta$ is block-wise diagonal, where each diagonal block takes the form of 
\[
\begin{pmatrix}1 & h\\ 
h & 1
\end{pmatrix}, \qquad \mbox{$h\in (-1,1)$ is a parameter}.     
\] 
The parameter $\sigma^2 = 0$ so there is no noise.  
The vector $\beta$ only has three nonzeros (but we don't know either the number of signals, or the locations or strengths of them):  
\[
\beta_1 = \tau,  \qquad  \beta_2  =\beta_3 = a \tau,  \qquad \mbox{where $\tau > 0$ and $a\in\mathbb{R}$}. 
\]  
In this simple setting, even there is no factors in $\Theta$, MR still does not perform well. 
%
For example, by direct calculations, 
\[
\beta_2 =a \tau, \quad  \beta_4 = 0,  \quad |(x_2,y)|/(x_2,x_2) =   |a - h| \tau,  \quad    |(x_4, y)|/(x_4,x_4)  = |a h| \tau.  
\] 
Therefore, we may face severe signal cancellation at location $2$, and variable $2$ (a signal variable) is ranked under variable $4$ (a noise variable) when 
\[
|a h| > |a - h|. 
\] 

We recognize that this problem can be resolved by exploiting local covariate structures. For each variable $j$, let 
\[
{\cal A}_j = \{\call_1, \call_2\}, \quad \call_1=\{j\}, \;\;\mbox{$\call_2=\{j,j+1\}$ for $j$ odd and $\call_2=\{j-1,j\}$ for $j$ even.}
\]
Each element $\call\in{\cal A}_j$ is called a ``neighborhood" of $j$. For each $\call\in{\cal A}_j$, we measure the
``significance" of variable $j$ in $\call$ by 
 \[
 T_{j | \call} = \|P_{\call} y\|^2 -
\|P_{{\cal I}\setminus \{j\}}y\|^2, 
\]
where $P_{{\cal I}}$ is the projection from $\mathbb{R}^n$ to the space spanned by $\{x_j, j \in {\cal I}\}$. Neglecting the influence of all variables outside the set ${\cal I}$, $T_{j|\call}$ is the 
likelihood ratio for testing whether $\text{supp}(\beta)=\call$ or $\text{supp}(\beta)={\cal I}\setminus\{j\}$.  
Take an odd $j$ for example, where $\call_1=\{j\}$ and $\call_2=\{j,j+1\}$. By direct calculations, 
\[
T_{j|\call_1}=n(\beta_j+h\beta_{j+1})^2, \qquad T_{j|\call_2}=n(1-h^2)\beta_j^2.
\]
When both variables $j$ and $(j+1)$ are signals, signal cancellation only affects $T_{j|\call_1}$ but not $T_{j|\call_2}$, so the latter is preferred. When variable $j$ is a signal and variable $(j+1)$ is a noise, signal cancellation affects neither of them; since $T_{j|\call_1}=n\beta_j^2\geq T_{j|\call_2}$ in this case, $T_{j|\call_1}$ is preferred. This motivates us to assess the significance of variable $j$ by combining these scores:
\[
T_j^* = \max \big\{T_{j|\call}: \call\in {\cal A}_j\big\}.
\]
In the above example,  
\[
T_2^* =n\max\bigl\{a^2(1-h^2), (a-h)^2 \bigr\}\tau^2, \qquad T_4^* = n(ah)^2\tau^2,   
\]   
and variables $2$ and $4$ are ranked correctly as long as $|h|<1/\sqrt{2}\approx 0.7$.

In more general cases, we use $\Theta$ to construct a graph and let a ``neighborhood" of $j$ be a connected subgraph that contains $j$, and the above idea can thus be conveniently extended. This is the main idea of the CAR step.

{\bf Remark}. It is possible to measure the significance differently by adding a cleaning step as in \cite{KJF-CASE},   
and hopefully we can evaluate variable $4$ with a score of $0$. However, cleaning usually requires a few tuning parameters, which is the first thing we wish to avoid  in variable ranking.  

{\bf Remark}. It is also possible to use the coefficients of a penalized-regression estimator  (e.g., the lasso) for ranking. Such estimators still require a critical tuning parameter that we wish to avoid. Additionally, when noise presents ($\sigma^2\neq 0$), these methods are less effective than screening methods for the blockwise-diagonal design; see \cite{JZZ-GS} and Section~\ref{subsec:SSsize}.

\subsection{Factor-adjusted Covariate Assisted Ranking (FA-CAR)} \label{subsec:method}
We extend the intuition gained in illustrating examples and develop a ranking method that works for a general design from Model \eqref{mod-G}. The method consists of a Factor-Adjusting (FA) step and a Covariate Assisted Ranking (CAR) step. 

In the FA step, let $X=
\sum_{k=1}^n \hat{\sigma}_k \hat{u}_k \hat{v}_k'$ be the SVD of $X$, where $\hat{\sigma}_1\geq
\hat{\sigma}_2\geq\cdots \geq \hat{\sigma}_n>0$ are the singular values, and
$\hat{u}_k\in\mathbb{R}^{n}$ and
$\hat{v}_k\in \mathbb{R}^{p}$ are the $k$-th (unit-norm) left and right
singular vectors, respectively. Introduce
\beq \label{tXty}
\tilde{y}=y-\sum_{k=1}^K
(\hat{u}_k'y)\hat{u}_k, \qquad \tilde{X}=X -\sum_{k=1}^K \hat{\sigma}_k\hat{u}_k\hat{v}_k'. 
\eeq
If we consider the two projection matrices $H_u=I_n - \sum_{k=1}^K \hat{u}_k\hat{u}_k'$ and $H_v=I_p - \sum_{k=1}^K \hat{v}_k\hat{v}_k'$, then it follows from elementary linear algebra that $\tilde{y}=H_uy$, $\tilde{X}=H_uX=XH_v$, and $(1/n)\tilde{X}'\tilde{X}=H_v\Theta H_v$.  As a result,
\beq \label{linearmod3}
\tilde{y} = \tilde{X}\beta + \tilde{z}, \quad \mbox{where } \tilde{z}\sim N(0,
\sigma^2 H_u ) \mbox{ and } n^{-1}\tilde{X}'\tilde{X}=\Theta - \sum_{k=1}^K (\hat{\sigma}_k^2/n)\hat{v}_k\hat{v}_k'. 
\eeq
This gives a new linear model with the same $\beta$ but a different Gram matrix. 
Note that $(\hat{\sigma}_k^2/n)$ and $\hat{v}_k$ are the $k$-th eigenvalue and eigenvector of $\Theta$, respectively. In Model \eqref{mod-G}, the component $G_0$ is a sparse matrix. Therefore, the leading eigenvalues (eigenvectors) of $\Theta$ are approximately equal to the leading eigenvalues (eigenvectors) of $(\Theta-G_0)$, i.e., $(\hat{\sigma}_k^2/n)\approx \lambda_k$ and $\hat{v}_k\approx v_k$ for $1\leq k\leq K$. We thus have 
\[
(1/n)\tilde{X}'\tilde{X}\approx \Theta - \sum_{k=1}^K\lambda_kv_kv_k'=G_0. 
\] 
So the Gram matrix for Model \eqref{linearmod3} is sparse.

In the CAR step, we focus on Model \eqref{linearmod3}. Write $\tilde{X}=[\tilde{x}_1,\ldots,\tilde{x}_p]$ and $G=(1/n)\tilde{X}'\tilde{X}$. Given a threshold $\delta\in (0,1)$, let $\cG^\delta$ be the graph with nodes $\{1,2,\ldots,n\}$ such that nodes $i$ and $j$ are connected by an undirected edge if
\beq \label{def-graph}
|G(i,j)|/\sqrt{G(i,i)G(j,j)}>\delta, \qquad 1\leq i\neq j\leq p.
\eeq
For each variable $j$, any connected subgraph $\call$ of $\cG^\delta$ that contains $j$ is called a ``neighborhood" of $j$. Consider a collection of such local ``neighborhoods"
\beq \label{Def-Aj}
{\cal A}_{\delta,j}(m)=\big\{\mbox{$\call$ is a connected subgraph of 
${\cal G}^\delta$}: j\in \call,\; 
|\call|\leq m  \big\}, 
\eeq
where $m\geq 1$ is an integer that controls the maximum size of selected neighborhoods. For $\call\in {\cal A}_{\delta,j}(m)$, we measure the ``significance" of variable $j$ in $\call$ by 
\beq \label{LocalScore} 
 T_{j | \call} = \|P_{\call} \tilde{y}\|^2 -
\|P_{{\cal I}\setminus \{j\}}\tilde{y}\|^2, 
\eeq
where $P_{{\cal I}}\tilde{y}$ is the projection of $\tilde{y}$ onto the space spanned by $\{\tilde{x}_j: j \in {\cal I}\}$. We then measure the ``significance" of variable $j$ by combining these scores:
\beq \label{Score}
T_j^* = \max \big\{T_{j|\call}: \call\in {\cal A}_{\delta,j}(m)\big\}.
\eeq
The scores $T_1^*,T_2^*,\ldots,T_p^*$ are used to rank variables.

FA-CAR has tuning parameters $(m,K,\delta)$, but the ideal choice of tuning parameters is insensitive to the unknown $\beta$ (it mainly depends on the design $X$). Therefore, tuning here is not as critical as it is for variable selection. In practice, we recommend using $m=2$ and $\delta=0.5$ and choosing $K$ as the elbow point in the scree plot of the Gram matrix $\Theta$ (see Section~\ref{sec:simu}).

The computational cost of our method comes from two parts: the SVD on $X$ and the CAR step. SVD is a rather manageable algorithm even for large matrices \citep{halko2011finding}. 
The computational cost of the CAR step is determined by the total number of subsets in the collection $\cA_\delta(m)\equiv \cup_{j=1}^p\cA_{\delta,j}(m)$. By graph theory \citep{frieze1999splitting},
\[
|{\cal A}_\delta(m)|\leq pm(2.72 d_p)^{m}, \qquad \mbox{$d_p$: maximum node degree}.
\]
Since 
$G\approx G_0$ is sparse, $d_p$ grows slowly with $p$. So the computational
cost of the CAR step is only moderately larger than that of MR.

Our method can also be conveniently extended to generalized linear models (GLM), which we discuss in Section~\ref{sec:GLM}.

\subsection{Application to a microarray dataset}  \label{subsec:realdata}
\begin{figure}[t]
\centering
\includegraphics[width=.32\textwidth, height=.23\textwidth]{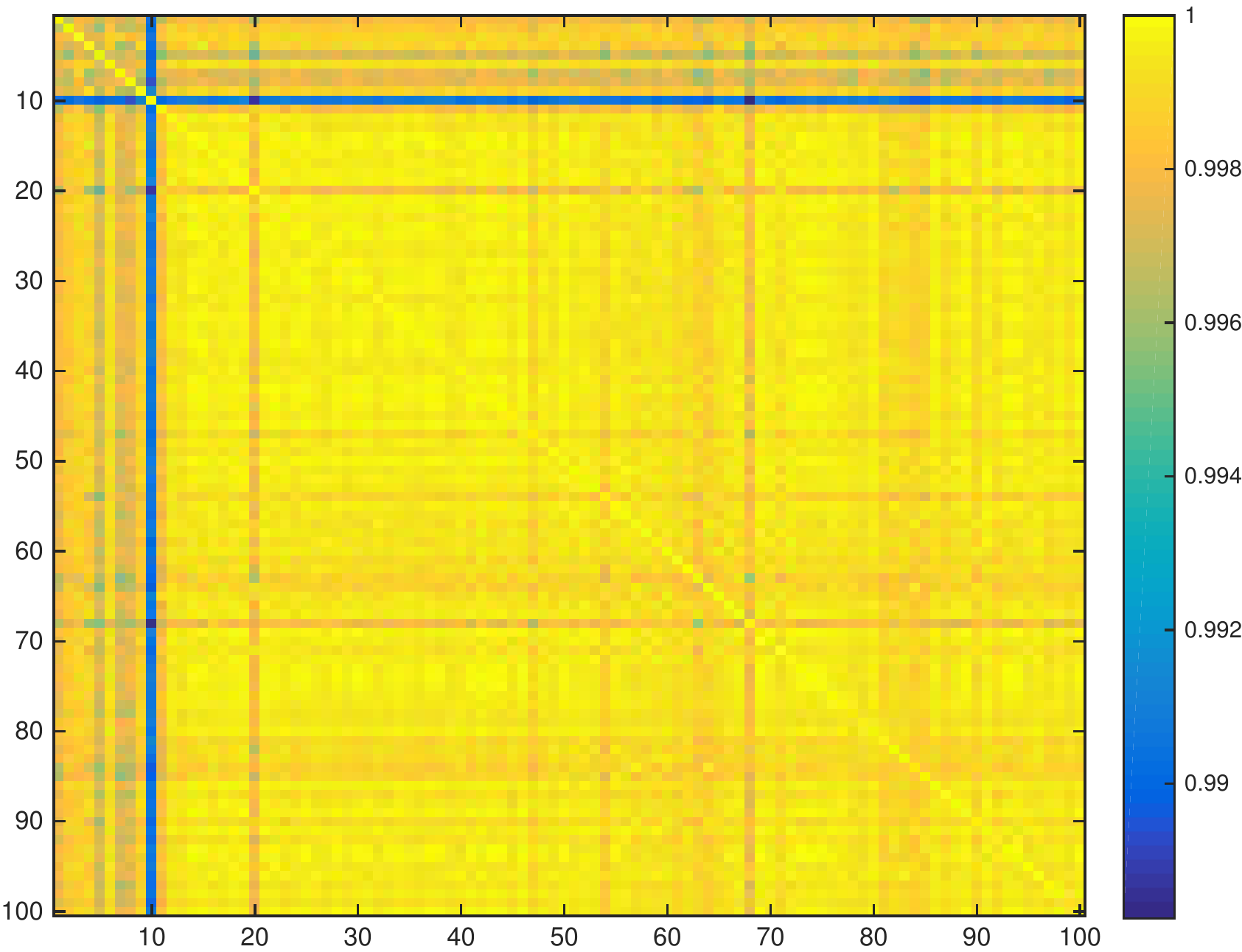}
\includegraphics[width=.31\textwidth, height=.23\textwidth]{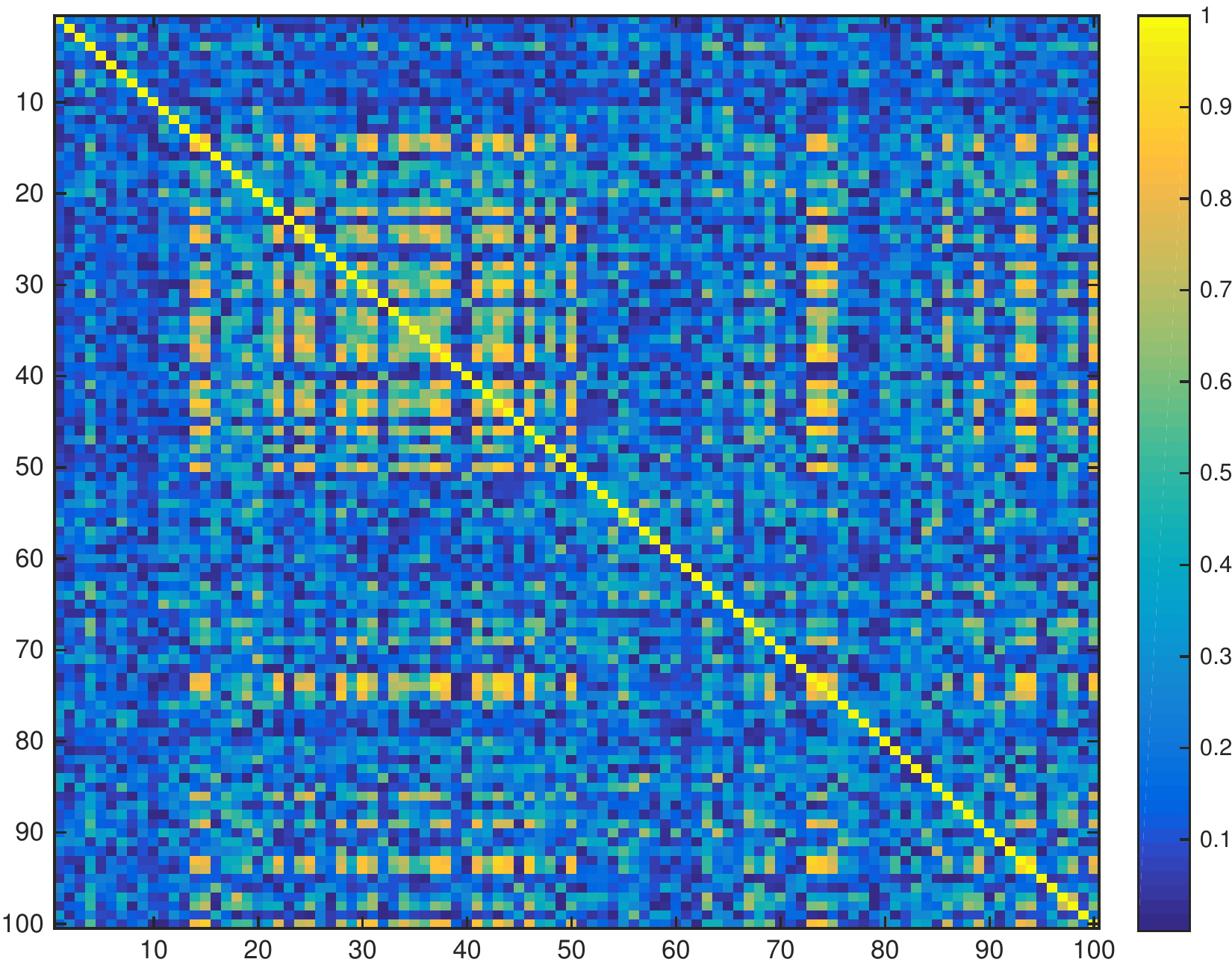}
\includegraphics[width=.31\textwidth, height=.23\textwidth]{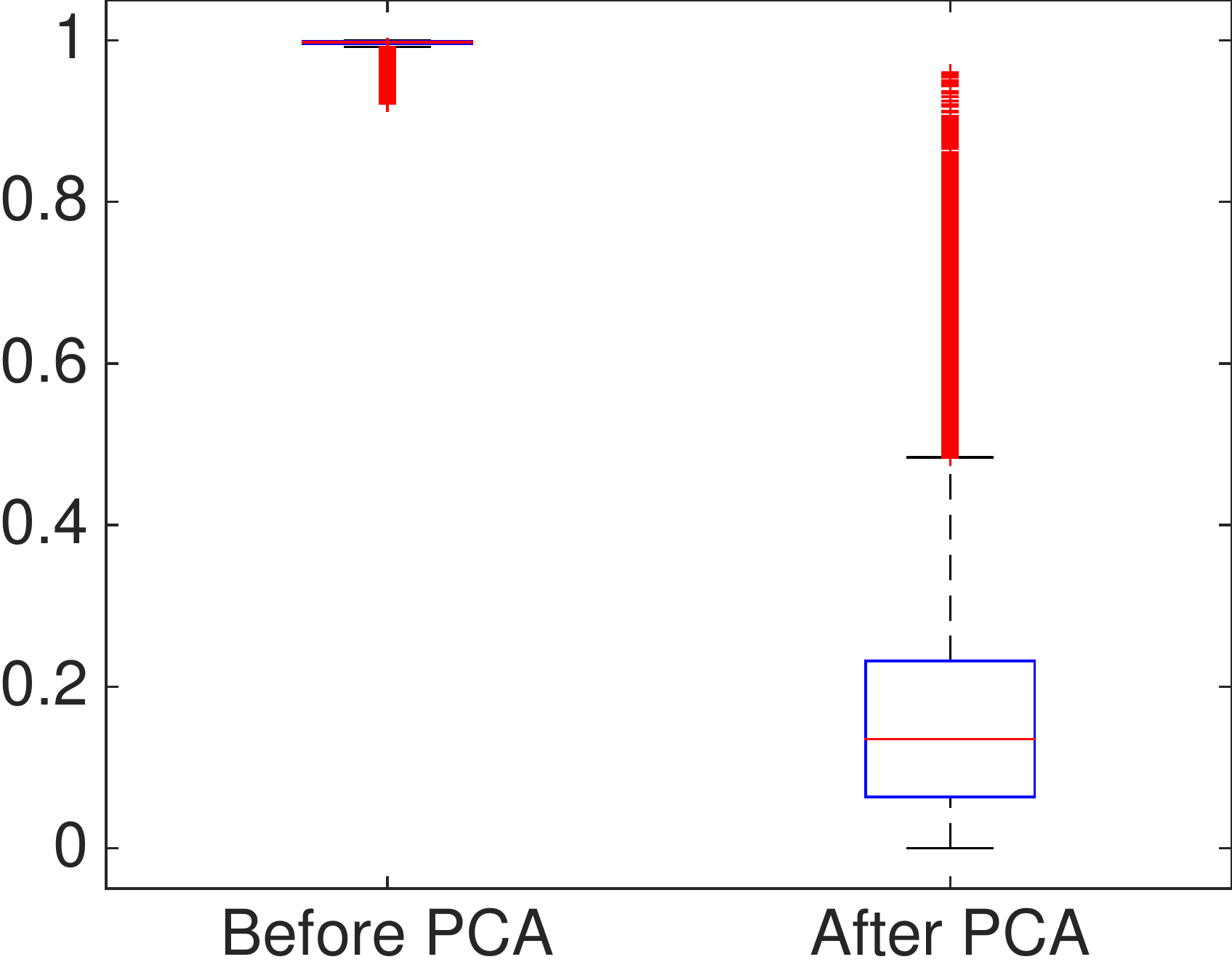}
\caption{Left two panels: the Gram matrix before and after Factor Adjusting (for presentation purpose, both matrices have been normalized so that the diagonals are $1$; only the upper left $100\times 100$ block is displayed). Right panel: Boxplots of the off-diagonal entries (in absolute value) of two Gram matrices.} \label{fig:nayak-Gram}
\end{figure}

We investigate the performance of our method using a gene microarray dataset \citep{nayak2009}. It contains the gene expressions of human immortalized B
cells for $p=4238$ genes and $n=148$ subjects (CEPH-Utah subpopulation). We use this $n\times p$ data matrix as the design. 
Figure~\ref{fig:nayak-Gram} compares the two respective Gram matrices for Model \eqref{linearmod} and Model \eqref{linearmod3}, and it shows that the Gram matrix for Model \eqref{linearmod3} is much sparser. This suggests that our assumption \eqref{mod-G} fits the data well and that the FA step is effective in removing the factors. 

We then compare our method with three other ranking methods on synthetic experiments. 
MR uses marginal correlation coefficients to rank variables. HOLP \citep{HOLP} and RRCS  \citep{li2012robust} are variants of MR: HOLP uses the absolute coordinates of $X'(XX')^{-1}y$ to rank variables, and RRSC uses the Kendall's $\tau$ marginal correlation coefficients. We measure the ranking performance using the Receiver Operating Characteristic (ROC) curve: Given the rank of variables, the ROC curve is obtained by successively retaining more variables. 

\begin{figure}[t]
\includegraphics[width=.315\textwidth]{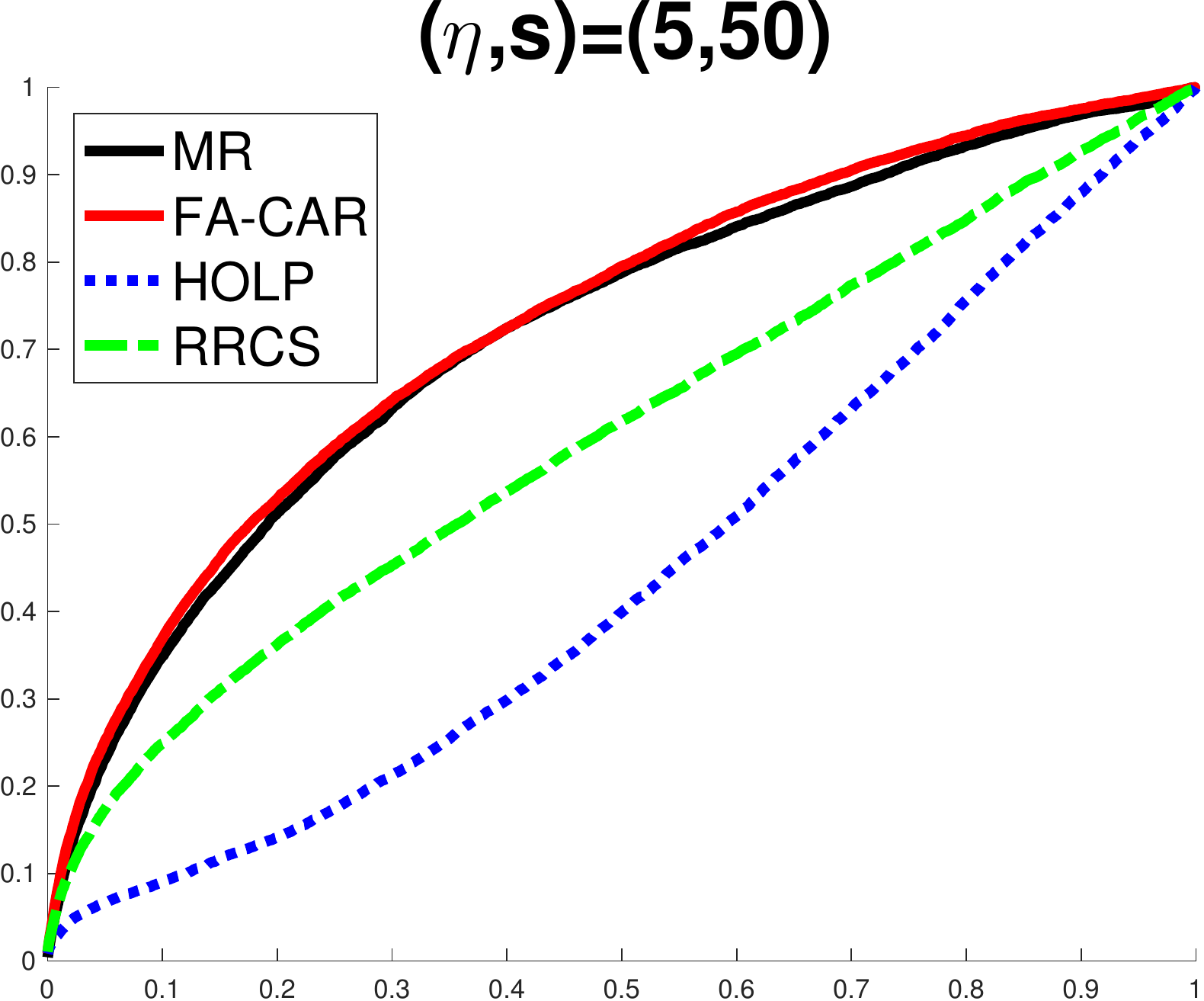}
\includegraphics[width=.315\textwidth]{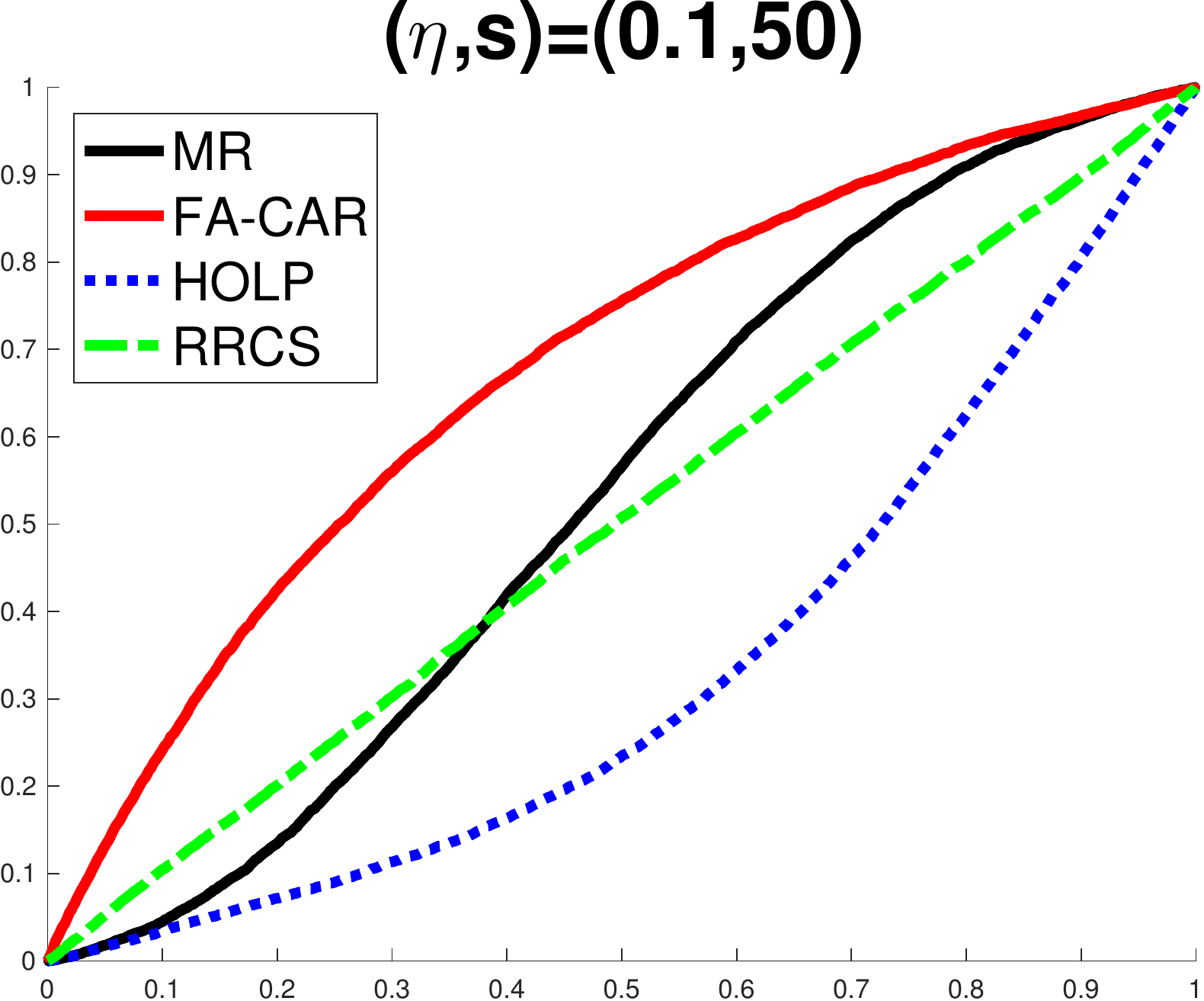}
\includegraphics[width=.315\textwidth]{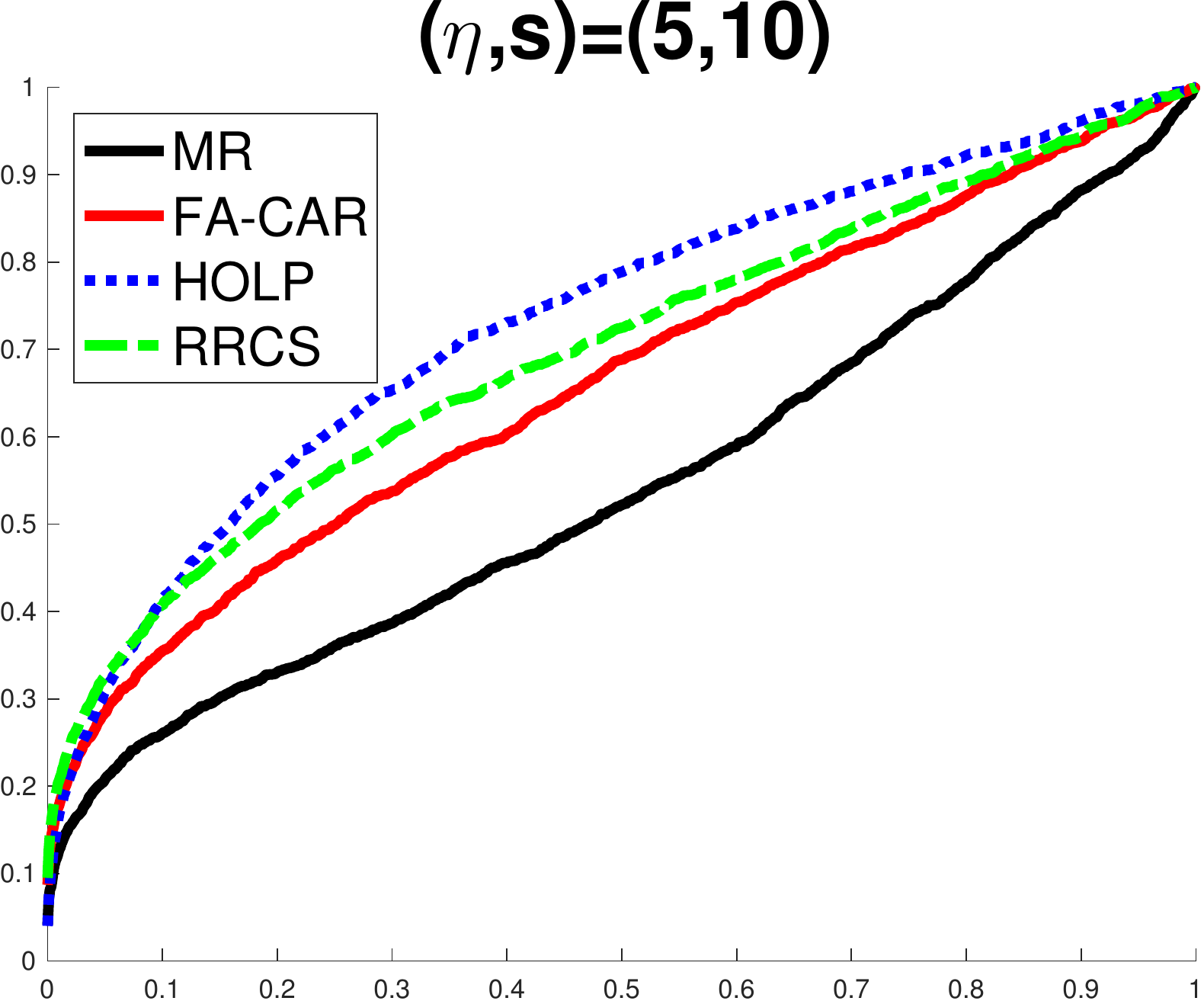}
\caption{The ROC. Design: gene-microarray. The curves are averaged over $200$ repetitions.}\label{fig:nayak-ROC}
\end{figure}

In our experiment, fixing parameters $(\eta,s)$, we first generate $\beta$ by drawing its first $s$ coordinates $iid$ from
$N(0,\eta^2)$ and setting the other coordinates to be $0$ and then
generate $y$ using Model \eqref{linearmod} with $\sigma=1$. Here, $(\eta,s)$ control the signal strength and signal sparsity, respectively. For each method, we report the average ROC curves over $200$ repetitions; the results are displayed in Figure~\ref{fig:nayak-ROC}. When $s=50$, FA-CAR always yields the best performance, and it is especially advantageous when $\eta$ is small (i.e., the signals are ``weak"). When $s=10$, FA-CAR performs reasonably well, and it is better than MR. It is a little worse than HOLP and RRSC, but these two methods are unsatisfactory in the other settings. In terms of the overall performance, we conclude that FA-CAR is the best among the four methods.

\subsection{Comparison of the sure-screening model size} \label{subsec:SSsize}
We use the blockwise-diagonal example in Section~\ref{subsec:idea} to demonstrate the advantage of exploiting local covariate structures for ranking. We use the sure-screening model size as the loss function, which is the minimum number of top-ranked variables one needs to select such that all signals are retained (then, all signal variables will be included in the follow-up lab experiments, say).   


We adopt a Rare/Weak (RW) signal model, which has been used a lot in the literature \citep{donoho2004higher, ji2012ups}. 
Fixing $\vartheta\in (0,1)$ and $r>0$, we assume the vector $\beta$ is generated from ($\nu_a$: a point mass at $a$)
\beq \label{simpleRW}
\beta_j\overset{iid}{\sim}(1-\eps_p)\nu_0 + \frac{\eps_p}{2}\nu_{\tau_p}+\frac{\eps_p}{2}\nu_{-\tau_p}, \quad \eps_p=p^{-\vartheta}, \tau_p=\sigma\sqrt{2r\log(p)/n}. 
\eeq
Under \eqref{simpleRW}, the total number of signals is approximately $s_p\equiv p^{1-\vartheta}$; as $p$ grows, the signals become increasingly sparser. The two parameters $(\vartheta,r)$ characterize the signal rareness and signal weakness, respectively. For any threshold $t>0$, let $FN_p(t)=\sum_{j=1}^p \mathbb{P}(\beta_j\neq 0, T_j^*\leq t)$ and $FP_p(t)=\sum_{j=1}^p \mathbb{P}(\beta_j=0, T_j^*>t)$ be the expected number of false negative and false positives, respectively. Define the sure-screening model size as 
\[
SS^*_p(\vartheta,r,h) = s_p  +  \min_{t: FN_p(t)<1} FP_p(t). 
\]

For the blockwise diagonal design, the Gram matrix is already sparse, so the FA step is not needed. We compare CAR with two other ideas, MR and LSR, where LSR simultaneously runs least-squares on each pair of variables $\{2j-1,2j\}$ for $j=1,2,\ldots,p/2$ and uses these least-squares coefficients to rank variables. 
We note that the least-squares estimator coincides with the recent de-biased lasso estimator \citep{ZZ14,vdGBRD14} in this design.
The following lemma shows that the convergence rate of $SS_p^*(\vartheta,r,h)$ for CAR is always no slower than those of the other two methods. 

\begin{lemma}[Sure-screening model size] \label{lem:block-rates}
Consider Model \eqref{linearmod} with the blockwise-diagonal design as in Section~\ref{subsec:idea}, where the RW model \eqref{simpleRW} holds. Let $L_p$ denote a generic multi-$\log(p)$ term such that $L_pp^{-\delta}\to 0$ and $L_pp^{\delta}\to\infty$ for all $\delta>0$. Given any  $(\vartheta, r, h)\in (0,1)\times (0,\infty)\times (-1,1)$, for each of the three methods, there is a constant $\eta^*(\vartheta,r,h)\in [0,1]$ such that $SS^*_p(\vartheta,r,h)=L_pp^{\eta^*(\vartheta,r,h)}$. 
Furthermore, for all $(\vartheta,r,h)$, 
\[
\eta^{*}_{CAR}(\vartheta,r,h)\leq \min\big\{ \eta^{*}_{MR}(\vartheta,r,h), \;\; \eta^{*}_{LSR}(\vartheta,r,h)\bigr\}. 
\]
\end{lemma}

The explicit expression of $\eta^*(\vartheta,r,h)$ for all three methods can be found in Lemma~\ref{lem:block-rates2}. Using the results there, we can find settings where the convergence rate of CAR is strictly faster; see Table~\ref{tb:block-rates}. 
\begin{table}[h]
\centering
\caption{The exponent $\eta^*(\vartheta,r,h)$ for the blockwise-diagonal design.} \label{tb:block-rates}
\begin{tabular}{c|c|c|c}
\hline
$(\vartheta, r, |h|)$ & $(.8, 1.5, .4)$ & $(.5, 2, .8)$ & $(.3, 2, .2)$\\
\hline
 CAR &  .395 & .500  & .700\\
MR & .395 & .920 & .751\\
LSR &  .543 & .980 & .700\\
\hline
\end{tabular}
\end{table}

{\bf Remark}. One might wonder why LSR is not the best method for ranking. This is a consequence of signal sparsity. When the signals are sparse, most of the $2$-by-$2$ blocks that contain signals have exactly one signal, so the least squares estimator is less efficient than the marginal estimator for estimating a signal $\beta_j$. 

\subsection{Connections}
Our method is related to the recent ideas of Graphlet Screening (GS) \citep{JZZ-GS} and Covariate Assisted Screening and Estimation (CASE) \citep{KJF-CASE}. These methods also use $\Theta$ to construct a graph and use local graphical structures to improve inference. However, our settings and goals are very different, and our method/theory can not be deduced from previous works: (a) GS and CASE are for variable selection and it is unclear how to use them for variable ranking. (b) GS and CASE have more stringent assumptions on the Gram matrix and do not work for the general designs considered in this paper.

Our FA step is related to the idea of using PCA to remove factor structures in multiple testing \citep{FHG-PFA} and covariance estimation \citep{FLM-poet2}, but our FA step is designed for linear models and is thus very different. \cite{wang2012factor} used PCA to improve marginal screening, which is similar to our FA step; however, their PCA approach is only justified for a random design that comes from an exact factor model, and their theory is insufficient for justifying our FA step.

Our work is related to the literatures on ranking differently expressed genes \citep{chen2007gene}. Common gene-ranking approaches (e.g., $p$-value, fold-change) are connected to the idea of Marginal Ranking. The key idea of our method is to exploit correlation structures among variables to improve MR, and an extension of our method (Section~\ref{sec:GLM}) can be potentially used for gene-ranking. On a high level, our work is also related to feature ranking problem in machine learning \citep{guyon2003introduction}, but most methods in these literatures (e.g., wrappers, filters) are algorithm-based and are not designed specifically for linear models.

\subsection{Content and notations}
The remaining of this paper is organized as follows. 
Section~\ref{sec:theory} contains asymptotic analysis, and  
Section~\ref{sec:simu} contains numerical results. Section~\ref{sec:GLM} provides an extension to generalized linear models. The discussions are in Section~\ref{sec:discuss}. Proofs are relegated to Section~\ref{sec:proofs}.

Throughout this paper, 
For positive sequences $\{a_n\}_{n=1}^\infty$ and
$\{b_n\}_{n=1}^\infty$, we write $a_n=o(b_n)$, $a_n=O(b_n)$ and
$a_n\lesssim b_n$, if $\lim_{n\to\infty}(a_n/b_n)=0$,
$\lim\sup_{n\to\infty}(a_n/b_n)<\infty$, $\max\{a_n-b_n,0\}=o(1)$,
respectively. Given $0\leq q\leq\infty$, for any vector $x$,
$\|x\|_q$ denotes the $L_q$-norm of $x$; for any $m\times n$
matrix $A$, $\|A\|_q$ denotes the matrix $L_q$-norm of $A$; when $q=2$, it coincides with the 
the spectral norm, and we omit the subscript $q$.  $\|A\|_F$ denotes the Frobenius
norm and $\|A\|_{\max}$ denotes the entrywise max norm. When $A$ is symmetric,
$\lambda_{\max}(A)$ and $\lambda_{\min}(A)$ denote the maximum and
minimum eigenvalues, respectively. For two sets $\call\subset
\{1,2,\dots,m\}$ and $\cJ\subset \{1,2,\dots,n\}$, $A^{\call,\cJ}$
denotes the submatrix of $A$ formed by restricting the rows and
columns of $A$ to sets $\call$ and $\cJ$. For a vector $x\in
\mathbb{R}^p$ and set $\call\subset \{1,2,\dots,p\}$, $x^{\call}$
denotes the sub-vector of $x$ formed by restricting coordinates to
set $\call$. 

\section{Asymptotic analysis} \label{sec:theory}
We describe the asymptotic settings in Section~\ref{subsec:condition} and present the main results in 
Section~\ref{subsec:mainThm}; our main results contain the rate of convergence of the sure-screening model size. 
Section~\ref{subsec:FA-thm} contains some new perturbation bounds for PCA; they are the key for studying Factor Adjusting and are also useful technical tools for other problems. Section~\ref{subsec:CARthm} contains the proof of the main result.

\subsection{Assumptions} \label{subsec:condition}
We assume $\Theta$ has unit diagonals without loss of generality. 
Let $S$ be the support of $\beta$ and let 
$s_p=|S|$. We assume  
\beq \label{cond-sparsity} 
\log(p)/n \goto 0, \qquad s_p \leq p^{1-\vartheta} \quad \mbox{for some
$\vartheta\in(0,1)$.} 
\eeq 
Under \eqref{cond-sparsity}, it is 
known that $n^{-1/2}\sqrt{\log(p)}$ is the minimax order of
signal strength for successful variable selection \citep{ji2012ups}.
We focus on
the most subtle region that nonzero $\beta_j$'s are constant
multiples of $n^{-1/2}\sqrt{\log(p)}$. Fixing a constant $r>0$
that calibrates the signal strength and a constant $a>0$, we assume for any $j\in S$,
\beq \label{cond-beta} 
\tau_p\leq |\beta_j|\leq a\tau_p, \qquad
\mbox{where}\quad \tau_p=n^{-1/2}\sigma\sqrt{2r\log(p)}. 
\eeq
Model \eqref{cond-sparsity}-\eqref{cond-beta} is a non-stochastic version of the Rare/Weak signal model in the literatures \citep{JK-RW}.

The Gram matrix $\Theta$ satisfies model \eqref{mod-G}. For any integer $1\leq m\leq p$ and matrix
$\Omega\in\mathbb{R}^{p,p}$, define $\nu_m^{\ast}(\Omega)$ as the minimum possible eigenvalue of any $m\times m$ principal submatrix of $\Omega$. Fixing
$\gamma\in(0,1)$, $c_0, C_0>0$ and an integer $g\geq 1$, we introduce a class of sparse covariance matrices: 
\[
{\cal M}_p(g, \gamma, c_0, C_0) = \Big\{ \Omega\in\mathbb{R}^{p,p} \mbox{ is p.s.d.}: \nu_g^{\ast}(\Omega) \geq c_0, \; \max_{1\leq i\leq p}\sum_{j=1}^p |\Omega(i,j)|^{\gamma} \leq C_0 \Big\}.
\]
Recall that $G_0$ is the sparse component in Model \eqref{mod-G}. We assume
\beq \label{cond-design}
G_0\in{\cal M}_p(g, \gamma, c_0, C_0), \quad \lambda_1\leq c_1 \lambda_K, \quad  \lambda_K /\max\{s_p,\log(p)\} \goto \infty,
\eeq
where $c_1>0$ is a constant. Fixing a constant $b>0$, let $\cG_0^{\delta}$ be the undirected graph whose nodes are $\{1,\cdots,p\}$ and there is an edge between nodes $i$ and $j$ if and only if
\[
|G_0(i,j)|/\sqrt{G_0(i,i)G_0(j,j)}> \delta_p, \qquad \mbox{where}\quad \delta_p=b/\log(p).
\]
This graph can be viewed as the ``oracle" graph, and the graph $\cG^\delta$ used in our method is an approximation to $\cG_0^\delta$.  Let
$\cG_{0,S}^\delta$ be the induced graph by restricting nodes to
$S$. We assume that for a positive integer
$\ell_0\leq g$,
\beq \label{cond-graphlet}
\mbox{each component of
$\cG_{0,S}^{\delta}$ consists of $\leq \ell_0$ nodes}.\footnote{A component is a subgraph in which any two nodes are connected to each other by a path, and which is connected to no additional nodes in the graph.}
\eeq
This is an assumption on the correlation structures among signal
variables. It implies
that the signal variables divide into many groups, each
consisting of $\leq \ell_0$ variables, such that signals in distinct groups are only weakly correlated after the factors are removed.

FA-CAR has tuning parameters $(K, m, \delta)$. We choose $K$
adaptively by 
\beq \label{choose-K} 
K = \hat{K}_p =
\max\big\{1\leq k\leq n: \hat{\sigma}^2_k> n\log(p) \big\}, 
\eeq
where $\hat{\sigma}_k$ is the $k$-th leading singular value of
$X$. We choose $(m,\delta)$ such that, for some constant
$C>0$, 
\beq \label{choose-delta+m} 
\ell_0\leq m \leq g, \qquad
1.01\delta_p\leq \delta \leq C\delta_p, 
\eeq 
where $\delta_p=b/\log(p)$, $g$ and $\ell_0$ are defined in
\eqref{cond-design} and \eqref{cond-graphlet} respectively.

\subsection{Main result: Sure-screening model size} \label{subsec:mainThm}
Given the scores $T_1^*,\ldots,T_p^*$, if we threshold them by
\beq \label{threshold}
t_p(q) = 2 q \sigma^2 \log(p), \qquad \mbox{$q>0$ is a constant},  
\eeq
the set of retained variables is $\hat{S}(q) = \hat{S}_p(q; X,y) =\{1\leq j\leq p: T^*_j>t_p(q)\}$. 
Recall that $S$ is the support of $\beta$ and $s_p=|S|$.
The (asymptotic) sure-screening model size is defined as
\beq \label{def-SSsize}
SS^*_p(\vartheta,r;X, \beta) = s_p + \min\Bigl\{ \mathbb{E}(|\hat{S}(q)\setminus S|): \mbox{$q$ satisfies} \lim_{p\to\infty }\mathbb{E}(|S\setminus\hat{S}(q)|)=0\Bigr\}.   
\eeq

To describe the asymptotic behavior of $SS_p^*$, we introduce the quantities $\omega_j(r,m)$, $1\leq j\leq p$, where $r$ calibrates the signal strength and $m\geq 1$ is a parameter in FA-CAR. 
By assumption \eqref{cond-graphlet}, the set of signal variables $S$ has the decomposition $S=\call_1\cup\call_2\cup\ldots\cup\call_M$, where nodes in each $\call_k$ form a component of $\cG^\delta_{0,S}$ and $\max_{1\leq k\leq M}|\call_k|\leq \ell_0$. Fix $j$. There exists a unique $\call_k$ which contains $j$. 
For any $\call\subset\call_k$, let $N=\call\setminus\{j\}$, $F=\call_k\setminus\call$, and $A_{j|\call}^0=G_0^{j,j}-G_0^{j,N}(G_0^{N,N})^{-1}G_0^{N,j}$. Define
\beq \label{omega_j_I}
\omega_{j|\call}(r; G_0,\beta,\delta_p) =  \frac{n A_{j|\call}^0}{2\sigma^2\log(p)} \left\{  \beta_j + (A_{j|\call}^0)^{-1}
[G_0^{j,F} - 
G_0^{j,N}(G_0^{N,N})^{-1}G_0^{N,F}]\beta^{F} \right\}^2. 
\eeq
For each $m\geq 1$, define
\beq \label{omega_j(r,m)}
\omega_j(r,m; G_0, \beta,\delta_p) = \max\left\{ \omega_{j|\call}(r;G_0,\beta,\delta_p): \call\in{\cal A}_{\delta,j}(m), \call\subset\call_k \right\}. 
\eeq
We notice that $\omega_j(r,m)$ is a monotone increasing function of $m$. 
The following definition is useful:
\begin{definition} \label{def:Lp}
$L_p$, as a positive sequence indexed by $p$, is called a multi-$\log(p)$ term if for any fixed $c>0$, $L_pp^c\goto\infty$ and $L_pp^{-c}\goto 0$ as $p\goto\infty$.
\end{definition}

The following theorem gives an upper bound for rate of convergence

\begin{theorem}[Sure-screening model size] \label{thm:SS}
Under Model \eqref{linearmod}-\eqref{mod-G}, suppose \eqref{cond-sparsity}-\eqref{cond-graphlet} hold for fixed $(\vartheta, r, g, \ell_0, \gamma, c_0, C_0, c_1, a, b)$ such that $g\geq \ell_0$, and suppose the tuning parameters $(K, m,\delta)$ satisfy
\eqref{choose-K}-\eqref{choose-delta+m}. Define the constant 
\[
q^*(\vartheta,r,m;G_0,\beta,\delta_p) =\inf\left\{ q\geq 0: \limsup_{p\to\infty} \frac{\log\Big(\sum_{j\in S}
p^{-[(\sqrt{\omega_j(r,m)}-\sqrt{q})_+]^2}\Big)}{\log(p)} > 0\right\}. 
\]
Then, as $p\goto\infty$, 
\[
SS_p^*(\vartheta,r; X, \beta)  \leq 
 L_p p^{1-\min\{ \vartheta,\; q^*(\vartheta,r,m)\}}. 
\]
\end{theorem}

We use Theorem~\ref{thm:SS} to draw some conclusions. First, we introduce a lower bound for the quantities $\omega_{j}(r,m)$. Fix $j$ and let $\call_k$ be the component of $\cG^\delta_{0,S}$ that contains $j$. Write $N=\call_k\setminus\{j\}$ and $A_{j|\call_k}^0=G_0^{j,j}-G_0^{j,N}(G_0^{N,N})^{-1}G_0^{N,j}$. Define 
\beq \label{omega_j_star}
\omega^*_j(r; G_0,\beta,
\delta_p)=\frac{n A_{j|\call_k}^0}{2\sigma^2\log(p)}\beta_j^2. 
\eeq
This quantity depends on $\beta$ only through $\beta_j$, so there should be no ``signal cancellation" involved in our method, as justified in the following corollary. 
\begin{cor}[No signal cancellation] \label{cor:no-cancellation}
Suppose the conditions of Theorem~\ref{thm:SS} hold. Let $c_0$ be the same as that in \eqref{cond-design}. Then, 
\[
SS_p^*(\vartheta,r; X, \beta)  \leq 
 L_p p^{1-\min\{ \vartheta,\;  [(\sqrt{c_0r}-\sqrt{1-\vartheta})_+]^2 \}}. 
\]
As a result, as long as $r$ is properly large, $SS_p^*\leq L_ps_p$. 
\end{cor}
Due to signal cancellation, no matter how large $r$ is, there still exist choices of the signs and locations of nonzero $\beta_j$'s such that MR ranks some signal variables strictly lower than many noise variables and that $SS_p^*\gg L_p s_p$. In contrast, Corollary~\ref{cor:no-cancellation} demonstrates that FA-CAR successfully overcomes the ``signal cancellation" issue.

Next, we compare FA-CAR with an alternative approach which applies MR after the FA step.\footnote{Since the Gram matrix for Model \eqref{linearmod3} has unequal diagonals, we first normalize the columns of $\tilde{X}$ to have the same $\ell^2$-norm and then apply MR.} 

\begin{cor}[Advantage over FA-MR] \label{cor:compare}
Suppose the conditions of Theorem~\ref{thm:SS} hold. Let $\widetilde{SS}_{p}^{*}(\vartheta,r; X, \beta)$ be the sure-screening model size for FA-MR. Then, 
\[
SS_p^{*}(\vartheta,r; X, \beta)\leq L_p\cdot  \widetilde{SS}_p^{*}(\vartheta,r; X, \beta).
\]
\end{cor}

Corollary~\ref{cor:compare} demonstrates that FA-CAR is always no worse than FA-MR. Additionally, we have seen examples in Section~\ref{subsec:SSsize} where FA-CAR is strictly better. This justifies the need of exploiting local covariate structures.

\subsection{Perturbation bounds for PCA} \label{subsec:FA-thm}
The success of the FA step relies on a tight bound for  
 $\|G- G_0\|_{\max}$. To bound this quantity, we need develop to new perturbation results for PCA. We can rewrite 
\[
G- G_0 = \sum_{k=1}^K(\hat{\sigma}^2_k/n)\hat{v}_k\hat{v}_k' - \sum_{k=1}^K\lambda_k v_k v_k', 
\] 
where $v_k$ and $\hat{v}_k$ are the $k$-th eigenvector of $(\Theta-G_0)$ and $G_0$, respectively. In the simplest case of $K=1$, the problem reduces to deriving a sharp bound for $\|\hat{v}_1-v_1\|_\infty$. Unfortunately, the standard tool of sine-theta theorem
\citep{davis1970rotation} only yields a bound for $\|\hat{v}_1-v_1\|$, which is often too loose if used as a bound for 
 $\|\hat{v}_1-v_1\|_\infty$. We need the following lemma:

 \begin{lemma}[Perturbation of leading eigenvector] \label{lem:Eigvec1}
Consider $\Theta=\lambda_1v_1v_1'+G_0$, where $\lambda_1>0$, $\|v_1\|=1$, and $G_0\in\mathbb{R}^{p,p}$ is symmetric. Let $\hat{v}_1$ be the leading eigenvector of $\Theta$.
If $3\|G_0\|_{\infty}\leq \lambda_1$, then
\[
\min\{\|\hat{v}_1-v_1\|_\infty, \|\hat{v}_1+v_1\|_\infty\} \leq 12\lambda_1^{-1}
\|G_0\|_{\infty}\|v_1\|_{\infty}.
\] 
\end{lemma}

We compare it with the sine-theta theorem, which gives that $\min\{\|\hat{v}_1-v_1\|_\infty, \|\hat{v}_1+v_1\|_\infty\} \leq \min\{\|\hat{v}_1-v_1\|, \|\hat{v}_1+v_1\|\}\leq C\lambda_1^{-1}\|G_0\|$. Consider a case where each row of $G_0$ has at most $d_p$ nonzero entries. Since $\|G_0\|_\infty\leq  d_p\|G_0\|$. $ \|v_1\|_{\infty}$, our bound is sharper if $d_p\|v_1\|_\infty=o(1)$. 


For the case $K>1$, the eigenvectors are generally not unique (unless all the eigenvalues are distinct from each other). It makes more sense to bound $\|\sum_{k=1}^K\hat{v}_k\hat{v}_k' - \sum_{k=1}^K v_kv_k'\|_{\max}$. We have the following theorem:

 \begin{theorem} \label{thm:Eigen}
Let $\Theta=\sum_{k=1}^K \lambda_kv_kv_k'+G_0$, where
$\lambda_1\geq \lambda_2\geq \cdots \lambda_K >0$, $v_1,\cdots,
v_K\in\mathbb{R}^p$ are unit-norm, mutually orthogonal vectors,
and $G_0\in\mathbb{R}^{p,p}$ is symmetric. For $1\leq k\leq K$,
let $(\hat{\lambda}_k, \hat{v}_k)$ be the $k$-th leading
eigenvalue and associated eigenvector of $\Theta$. Write
$V=[v_1,\cdots, v_K]$, $\hat{V}=[\hat{v}_1,\cdots,\hat{v}_K]$ and
$G=\Theta - \sum_{k=1}^K\hat{\lambda}_k\hat{v}_k\hat{v}_k'$. If
$\lambda_K>C_1\|G_0\|_{\infty}$ for some constant $C_1>2$, then
\[
\|VV' - \hat{V}\hat{V}'\|_{\max}\leq C_2 (\lambda_1/\lambda_K)^2 \cdot \lambda_K^{-1} \|G_0\|_\infty\cdot \max_{1\leq k\leq
K}\|v_k\|_{\infty}^2,
\]
and
\[
\|G - G_0\|_{\max} \leq C_2'
(\lambda_1/\lambda_K)^2\cdot \|G_0\|_{\infty}\cdot \max_{1\leq k\leq
K}\|v_k\|_{\infty}^2,
\]
where $C_2,C_2'>0$ are constants that only depend on $(C_1, K)$. 
\end{theorem}

The proof of Lemma~\ref{lem:Eigvec1} uses a similar approach as the proof of Lemma 3.1 in \cite{jin2015phase}. The proof of Theorem~\ref{thm:Eigen} is new and highly non-trivial since we do not assume any gap between $\lambda_1,\ldots,\lambda_K$. That it requires no eigen-gaps makes this result very different from other recent perturbation results (e.g., \cite{ke2017new}).

\subsection{Proof of Theorem~\ref{thm:SS}} \label{subsec:CARthm}
Recall that $G=(1/n)\tilde{X}'\tilde{X}$ is the Gram matrix of Model \eqref{linearmod3}. Using the results in Section~\ref{subsec:FA-thm}, we can show that $G$ is entry-wise close to $G_0$:
\begin{lemma} \label{lem:FA}
Suppose the conditions of Theorem~\ref{thm:SS} hold. Then, 
$\|G-G_0\|_{\max} = o(\delta_p)$ and $s_p \|G-G_0\|_{\max} =o(1)$.  
\end{lemma}

The key of the proof is to study the distribution of $T_{j|\call}$, for each $j\in S$ and $\call\in{\cal A}_{\delta,j}(m)$. The following lemma is proved in Section~\ref{sec:proofs}. 
\begin{lemma} \label{lem:statistics}
Suppose the conditions of Theorem~\ref{thm:SS} hold. Fix $j\in S$ and let $\call^{(j)}$ be the unique component of $\cG_{0,S}^\delta$ that contains $j$. For any $\call\subset \call^{(j)} \cap \cG_{S}^\delta$ that contains $j$, 
\[
T_{j|\call} = (W +\Delta)^2, \quad W\sim \mathcal{N}\left(\sqrt{2\omega_{j|\call}(r)\sigma^2\log(p)},\; \sigma^2\right), \;\; |\Delta|=o_P\left(\sqrt{\log(p)}\right),  
\]
where $\omega_{j|\call}(r)$ is the same as that in Section~\ref{subsec:mainThm}. 
\end{lemma}

The proof of Lemma~\ref{lem:statistics} is lengthy, and we provide some illustration. For simplicity, we only consider a special case where $\call$ is exactly the component of $\cG_S^\delta$ that contains $j$.
By definition and elementary calculations (see \eqref{Tjcall-expression}), 
\beq \label{Tjcall-illust}
T_{j|\call} = n^{-1}(\eta^{\call})'\left( (G^{\call,\call})^{-1} -\begin{bmatrix}
(G^{N,N})^{-1} & 0\\ 0& 0\end{bmatrix}
\right) \eta^{\call}, 
 \quad \eta=\tilde{X}'\tilde{y}, \; N=\call\setminus\{j\},  
\eeq
Since $\eta\sim {\cal N}(nG\beta, \sigma^2 nG)$, we have 
\[
n^{-1}\mathbb{E}[\eta^{\call}] = (G\beta)^\call = G^{\call,\call}\beta^{\call} + G_0^{\call,\call^c}\beta^{\call^c} + (G-G_0)^{\call,\call^c}\beta^{\call^c}.  
\]
It can be proved that the third term is negligible as a result of Lemma~\ref{lem:FA} and the second term is negligible due to the sparsity of $G_0$ and the definition of the graph $\cG^\delta$. It follows that $E[\eta^{\call}]\approx n G^{\call,\call}\beta^{\call}$. We plug it into \eqref{Tjcall-illust} and find that  
\[
T_{j|\call}  \approx n (\beta^{\call})'\bigl[ (G^{\call,\call})^{-1}-G^{\call,N}(G^{N,N})^{-1}G^{N,\call}\bigr]\beta^{\call} = nA_{j|\call}\beta_j^2, 
\]
where $A_{j|\call}= G^{j,j}-G^{j,N}(G^{N,N})^{-1}G^{N,j}$ is a counterpart of $A_{j|\call_k}^0$ in \eqref{omega_j_star} and the last equality is a result of the matrix inverse formula in linear algebra. It remains to characterize the difference between $A_{j|\call}$ and $A_{j|\call_k}^0$; recall that $\call_k$ is the unique component of $\cG^\delta_{0,S}$ that contains $j$. Using Lemma~\ref{lem:FA}, we can prove that, if we restrict $\cG_{0,S}^\delta$ to $\call_k$, it splits into a few components and one component is exactly $\call$. Such an observation allows us to show that  
\[
A_{j|\call}=A^0_{j|\call_k}[1+o(1)]. 
\]
The proof of Lemma~\ref{lem:statistics} follows a similar idea as the above derivation but is much more complicated.

Once we have the distribution of $T_{j|\call}$, we can quantify the type I and type II errors associated with any threshold $t_p(q)$.

\begin{lemma}[Type I and Type II errors] \label{thm:typeI+II}
Suppose the conditions of Theorem~\ref{thm:SS} hold. Consider $\hat{S}(q)$, the set of selected variables associated with the threshold $t_p(q)$ as in \eqref{threshold}. Then,
\[
\mathbb{E}( |S\backslash \hat{S}(q)|) \leq L_p \sum_{j\in S}
p^{-[(\sqrt{\omega_j(r,m)}-\sqrt{q})_+]^2},
\]
and
\[
\mathbb{E}( |\hat{S}(q)\backslash S|) \leq C s_p [\log(p)]^{\gamma m} + L_p
p^{1-q}.
\]
where $\omega_j(r,m)$ is as in \eqref{omega_j(r,m)} and $\gamma$ is the same as that in ${\cal M}_p(g,\gamma,c_0,C_0)$. 
\end{lemma}

We now derive the upper bound for $SS_p^*(\vartheta,r;X,\beta)$. 
By Lemma~\ref{thm:typeI+II} and the definition of $q^*(\vartheta,r,m)$, for any $q<q^*(\vartheta,r,m)$, there is an $\epsilon>0$ such that 
\[
\mathbb{E}( |S\backslash \hat{S}(q)|) \leq L_pp^{-\eps}\goto 0, \qquad\mbox{for all sufficiently large $p$}. 
\]
As a result, for any $q<q^*(\vartheta,r,m)$, 
\[
SS_p^*(\vartheta,r;X,\beta)\leq s_p + \mathbb{E}( |\hat{S}(q)\backslash S|)\leq L_pp^{1-\min\{\vartheta,q\}}. 
\]
Taking the limit of $q\to q^*(\vartheta,r,m)$ gives the claim of Theorem~\ref{thm:SS}.

\section{Simulations}  \label{sec:simu}
We investigate the performance of FA-CAR in simulations. In all experiments below, given a covariance matrix
$\Sigma\in\mathbb{R}^{p,p}$, the rows of $X$ are independently
sampled from $\mathcal{N}(0,\Sigma)$. We consider four different
types of designs where the corresponding $\Sigma$ is:
\begin{itemize}
\item {\it Tridiagonal.} $\Sigma(j,j)=1$ for all $1\leq j\leq p$, and $\Sigma(i,j)=\rho \cdot 1\{|i-j|=1\}$ for any $1\leq i\neq j\leq p$. We set $\rho=0.5$.
\item {\it Autoregressive.}
$\Sigma(i,j)=\rho^{|i-j|}$ for all $1\leq i, j\leq p$. We set $\rho=0.6$.
\item {\it Equal correlation.} $\Sigma(j,j)=1$ for $1\leq j\leq p$, and $\Sigma(i,j)=\rho$ for $1\leq i\neq j\leq p$. We set $\rho=0.6$.
\item{\it Two factors.} $\Sigma = \frac{\rho}{2} a_1a_1' + \frac{\rho}{2} a_2a_2' +
(1-\rho)\Sigma_1$,
where $a_1=(1,1,\dots,1)'$, $a_2=(1,-1,1,-1,\dots,1,-1)'$ and
$\Sigma_1$ is an autoregressive covariance matrix, i.e.,
$\Sigma_1(i,j) = \rho_1^{|i-j|}$. We set $\rho=0.5$ and $\rho_1=0.6$.
\end{itemize}
Fixing $(n,p,\eta,s)$, we generate $\beta$ as
follows: The first $s$ coordinates of $\beta$ are independently
sampled from $\mathcal{N}(0,\eta^2)$, the other coordinates all
equal to 0. We then generate $y$ using Model \eqref{linearmod} with $\sigma^2=1$. 


Our method has three tuning parameters $(K, \delta, m)$. In Experiments 1-2, we set $m=2$ and $\delta=0.5$, and use the ideal choice of $K$, that is, $K=0, 0, 1,
2$ for the above four types of designs. In Experiment 3,  we investigate the sensitivity of our method to tuning parameters.

\begin{figure}[t]
\begin{tabular}{cc}
\includegraphics[width=0.45\textwidth,height=0.35\textwidth]{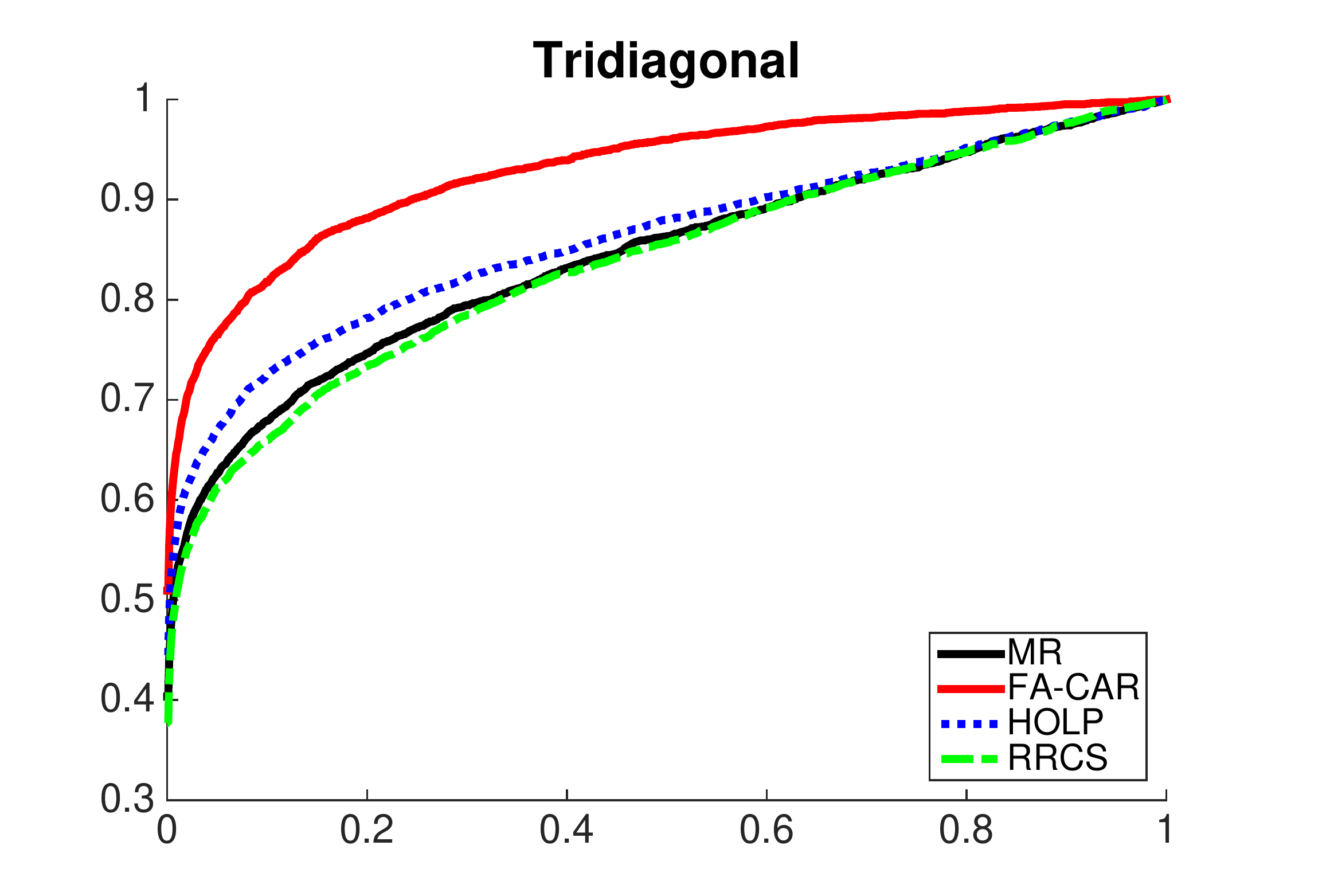} &
\includegraphics[width=0.45\textwidth,height=0.35\textwidth]{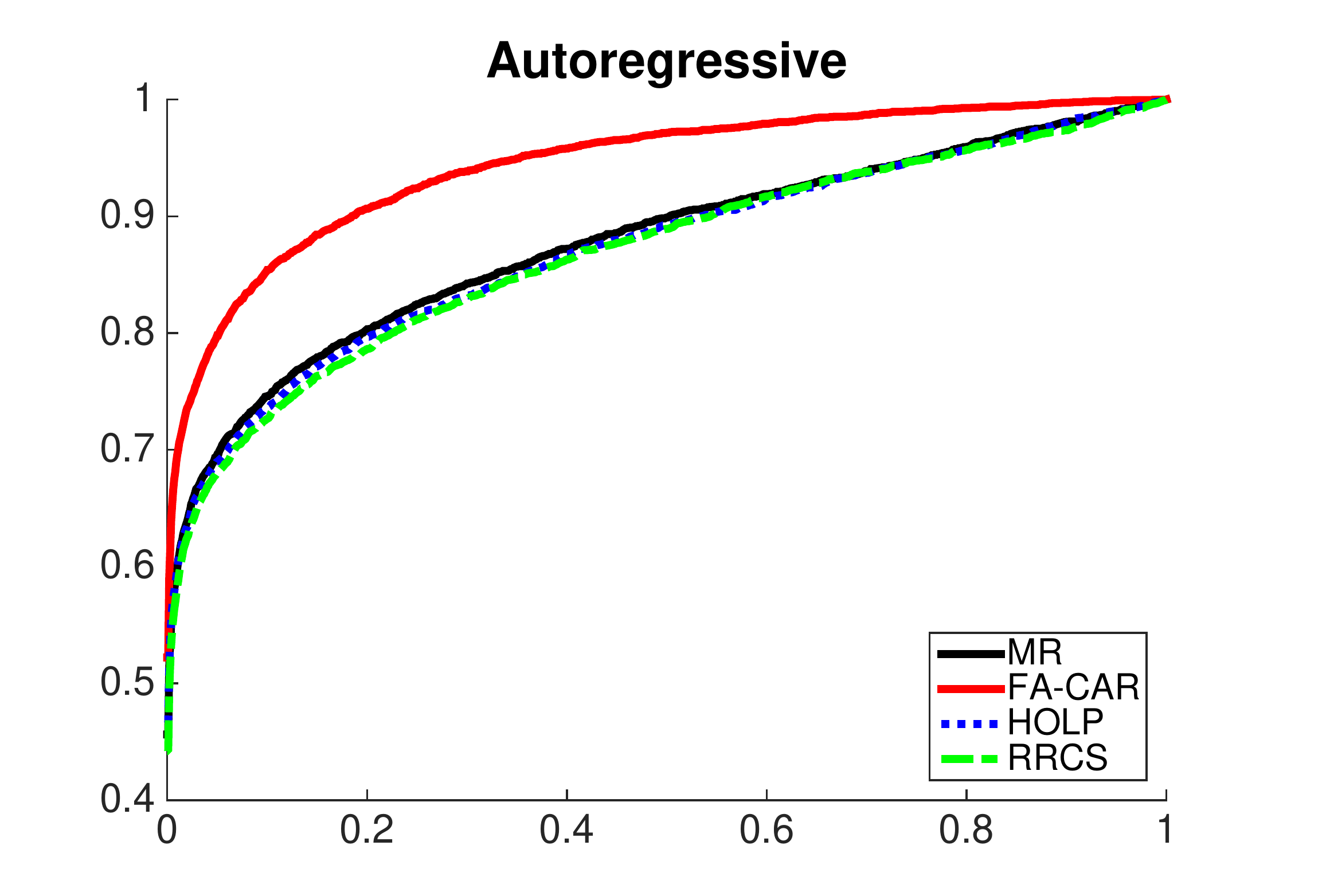}
\\
\includegraphics[width=0.45\textwidth,height=0.35\textwidth]{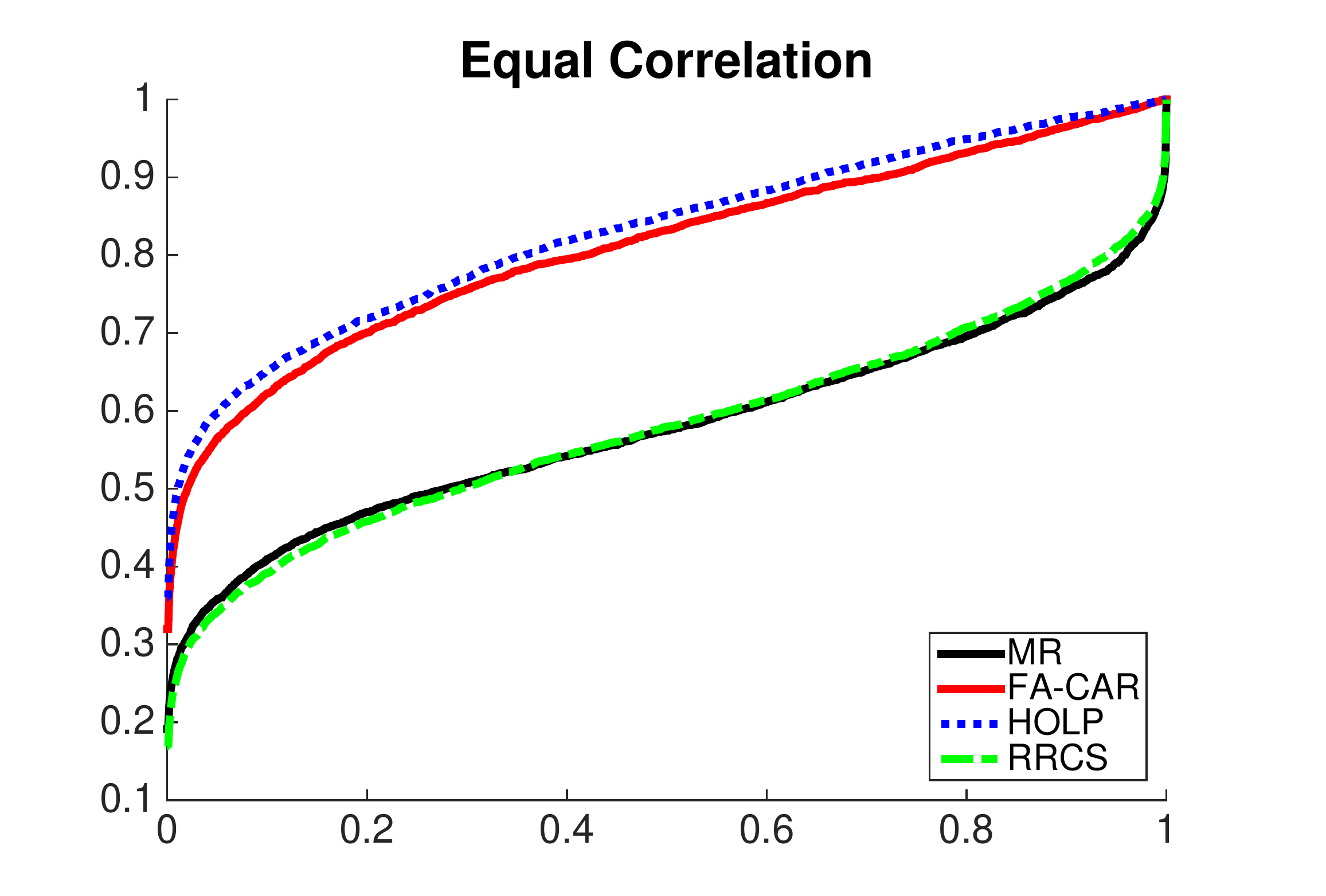} &
\includegraphics[width=0.45\textwidth,height=0.35\textwidth]{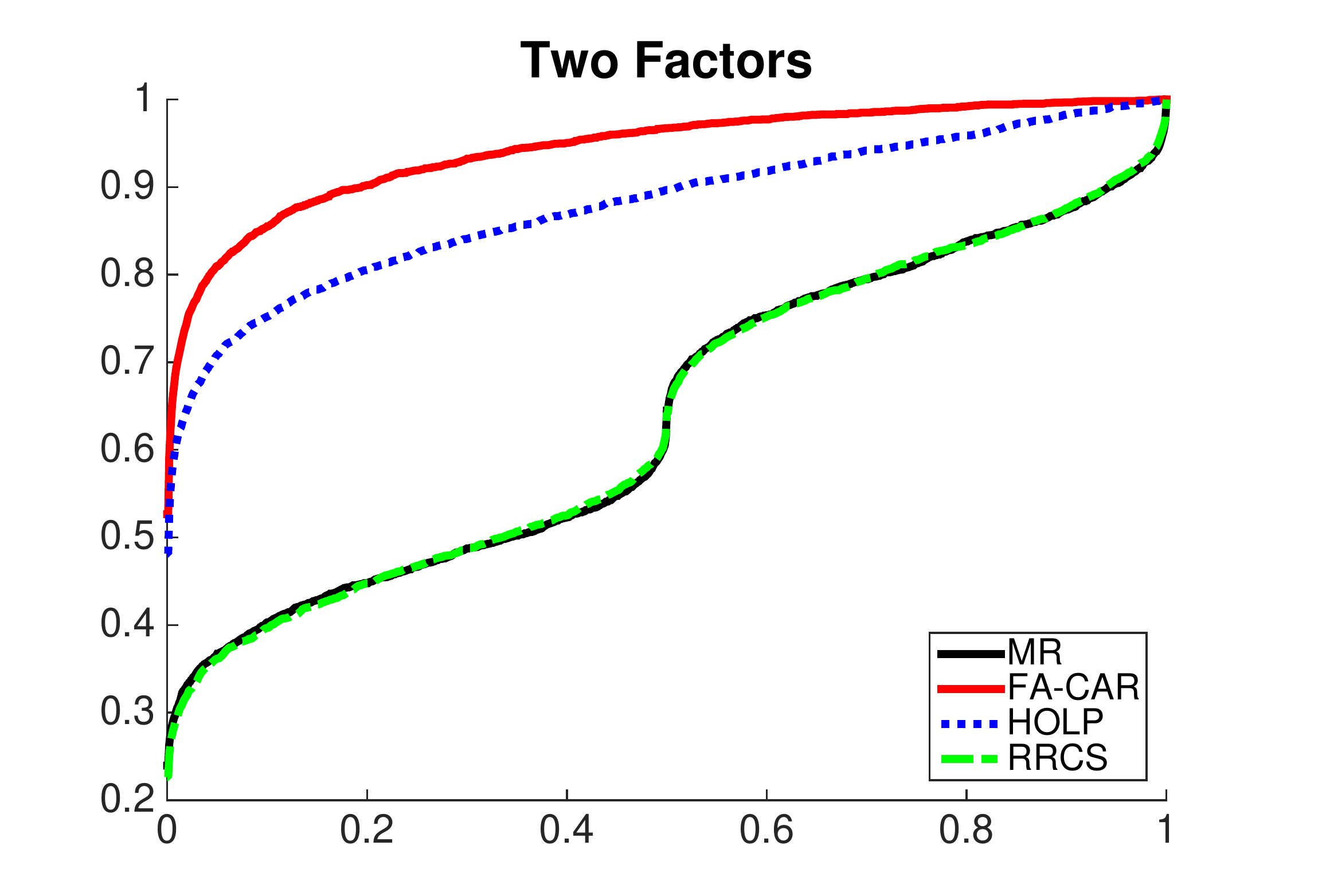}
\\
\end{tabular}
\caption{ROC curves in Experiment 1. $(n,p,\eta,s)=(200,1000,3,20)$.}\label{ROClinear}
\end{figure}

\paragraph{Experiment 1: Comparison of ROC curves} 
We compare the ROC curves of our method and three other methods:  (1) Marginal Ranking (MR)
\citep{FanLv-SS}, (2) HOLP \citep{HOLP}, which uses the coordinates of $X'(XX')^{-1}y$ for ranking, and (3) RRCS \citep{li2012robust} which uses the marginal Kendall's $\tau$ correlation coefficients for ranking.
Fix $(n,p,\eta,s)=(200,1000,3,20)$. For each of the four design types, we generate $200$ datasets and output the average ROC curves of these $200$ repetitions. The results are displayed in Figure~\ref{ROClinear}. For the tridiagonal, autoregressive, and two-factor designs, our method significantly outperforms the other methods. For the equal correlation design, our method significantly outperforms MR and RRCS and is similar to HOLP.

\paragraph{Experiment 2: Various $(n,p,\eta,s)$}
We consider four choices of $(n,p,\eta,s)$, for each of the four types of designs. 
There are 
 $16$ different settings in total.
 We measure the performance of different methods using several criteria: (a) Sure screening probability (SP): the probability that all the signal variables are selected when retaining $n$ variables in total.  (b) Type II: the number of type II errors when retaining $n$ variables in total. (c) Sure screening model size (Size):  the minimum number $L$ such that all signal variables are selected when retaining $L$ variables in total. 
The results are shown in Table~\ref{lineartable}.

\begin{table}[h]
\center \caption{Results of Experiment 2. For Type
II, we report the mean over $200$ repetitions, and for Size, we report
the median over $200$ repetitions.} \resizebox{1\columnwidth}{!}{
\begin{tabular}{c|c|c|c c c c}
\hline
\multirow{2}{*}{Designs}  & \multirow{2}{*}{Setting: $(n,p,\eta,s)$} & \multirow{2}{*}{Measure} & \multicolumn{4}{c}{Method} \\
\cline{4-7}
& & & FA-CAR & MR & HOLP & RRCS \\
\hline \multirow{8}{*}{Tridiag.} & \multirow{2}{*}{(200,1000,3,5)} & SP/Type II  & \textbf{0.91/0.11} & 0.45/0.64 & 0.51/0.58 & 0.45/0.65 \\
& & Size & \textbf{6} & 246 & 195 & 247 \\
 \cline{2-7} & \multirow{2}{*}{(200,1000,3,20)} & SP/Type II &
\textbf{0.11/2.45} & 0.01/5.19 & 0.01/4.49 & 0.01/5.47 \\
 & & Size & \textbf{518.5} & 865.5 & 861 & 874.5 \\
\cline{2-7} & \multirow{2}{*}{(200,1000,0.5,5)} & SP/Type II &
\textbf{0.73/0.35} & 0.38/0.86 & 0.36/0.94 & 0.35/0.95 \\
& & Size & \textbf{39.5} & 336.5 & 384.5 & 382 \\
\cline{2-7} & \multirow{2}{*}{(200,5000,0.5,5)} & SP/Type II &
\textbf{0.57/0.66} & 0.20/1.39 & 0.18/1.42 & 0.17/1.47 \\
 & & Size & \textbf{132.5} & 1789.5 & 1942 & 1964 \\
\hline
\multirow{8}{*}{Autoreg.} & \multirow{2}{*}{(200,1000,3,5)} & SP/Type II  & \textbf{0.95/0.06} & 0.67/0.43 & 0.62/0.46 & 0.67/0.44 \\
 & & Size & \textbf{6} & 65 & 85 & 58.5 \\
\cline{2-7} & \multirow{2}{*}{(200,1000,3,20)} & SP/Type II &
\textbf{0.17/2.00} & 0.02/4.05 & 0.00/4.19 & 0.01/4.37 \\
 & & Size & \textbf{422} & 840 & 848 & 850.5 \\
\cline{2-7} & \multirow{2}{*}{(200,1000,0.5,5)} & SP/Type II &
\textbf{0.79/0.28} & 0.53/0.67 & 0.42/0.83 & 0.51/0.77 \\
& & Size & \textbf{22} & 179.5 & 293 & 184 \\
\cline{2-7} & \multirow{2}{*}{(200,5000,0.5,5)} & SP/Type II &
\textbf{0.6/0.61} & 0.35/1.18 & 0.35/1.20 & 0.315/1.29 \\
& & Size & \textbf{57} & 937 & 980 & 1077 \\
\hline
\multirow{8}{*}{Equal corr.} & \multirow{2}{*}{(200,1000,3,5)} & SP/Type II  & 0.46/0.72 & 0.06/1.85 & \textbf{0.46/0.68} & 0.07/1.84 \\
 & & Size & 247 & 998 & \textbf{230} & 997 \\
\cline{2-7} & \multirow{2}{*}{(200,1000,3,20)} & SP/Type II &
0.00/6.09 & 0.00/10.66 & \textbf{0.00/5.73} & 0.00/10.86 \\
& & Size & 909.5 & 1000 & \textbf{863.5} & 1000 \\
\cline{2-7} & \multirow{2}{*}{(200,1000,0.5,5)} & SP/Type II &
0.16/\textbf{1.38} & 0.05/2.06 & \textbf{0.17}/1.47 & 0.03/2.08 \\
 & & Size & \textbf{577} & 969 & 589 & 957 \\
\cline{2-7} & \multirow{2}{*}{(200,5000,0.5,5)} & SP/Type II &
\textbf{0.07/2.00} & 0.00/2.65 & 0.06/2.03 & 0.00/2.72\\
 & & Size & \textbf{2600.5} & 4856 & 2689.5 & 4844.5 \\
\hline
\multirow{8}{*}{Two factors} & \multirow{2}{*}{(200,1000,3,5)} & SP/Type II  & \textbf{0.93/0.09} & 0.16/1.83 & 0.62/0.47 & 0.17/1.82 \\
 & & Size & \textbf{6} & 690.5 & 88.5 & 687.5 \\
\cline{2-7} & \multirow{2}{*}{(200,1000,3,20)} & SP/Type II &
\textbf{0.21/2.03} & 0.01/11.06 & 0.02/3.99 & 0.01/11.08 \\
 & & Size & \textbf{454} & 988.5 & 832.5 & 983.5 \\
\cline{2-7} & \multirow{2}{*}{(200,1000,0.5,5)} & SP/Type II &
\textbf{0.73/0.43} & 0.17/1.94 & 0.38/1.06 & 0.16/1.95 \\
 & & Size & \textbf{43.5} & 674.5 & 387 & 678 \\
\cline{2-7} & \multirow{2}{*}{(200,5000,0.5,5)} & SP/Type II &
\textbf{0.47/0.89} & 0.08/2.46 & 0.28/1.45 & 0.07/2.49 \\
& & Size & \textbf{274} & 3592.5 & 1382 & 3633 \\
\hline
\end{tabular}
}\label{lineartable}
\end{table}

\begin{figure}[t]
\includegraphics[width=0.45\textwidth,height=0.35\textwidth]{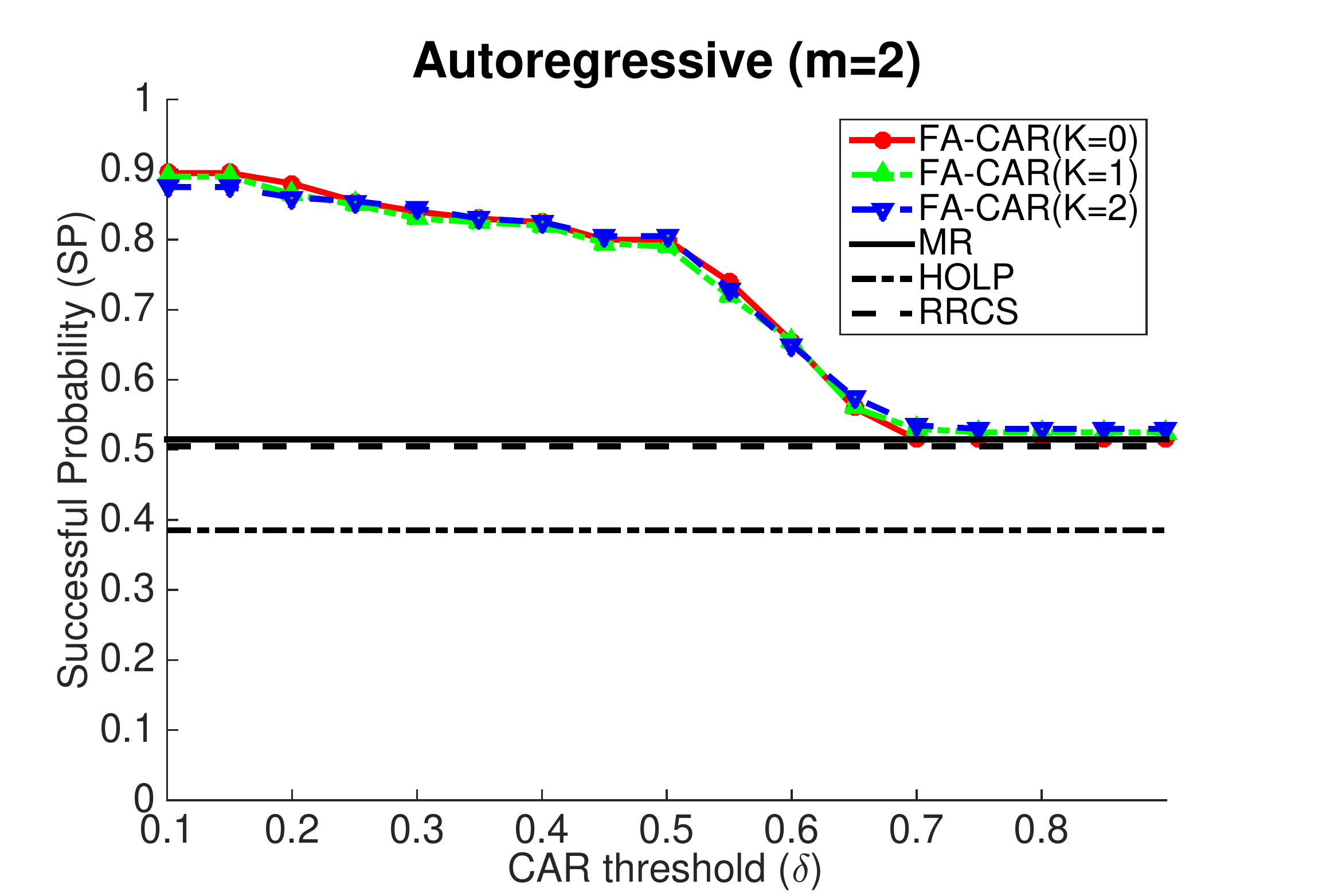} 
\includegraphics[width=0.45\textwidth,height=0.35\textwidth]{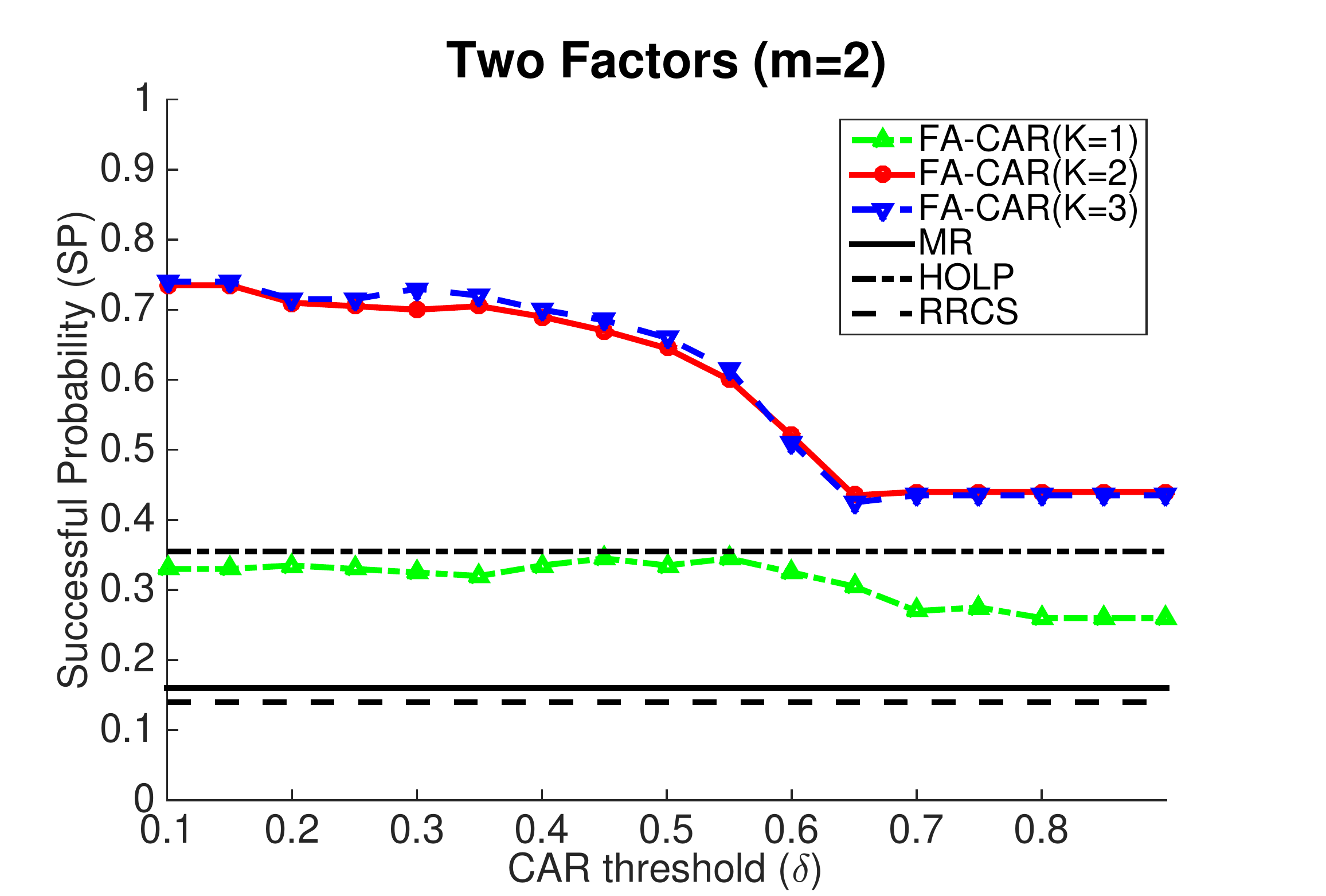} \\
\includegraphics[width=0.45\textwidth,height=0.35\textwidth]{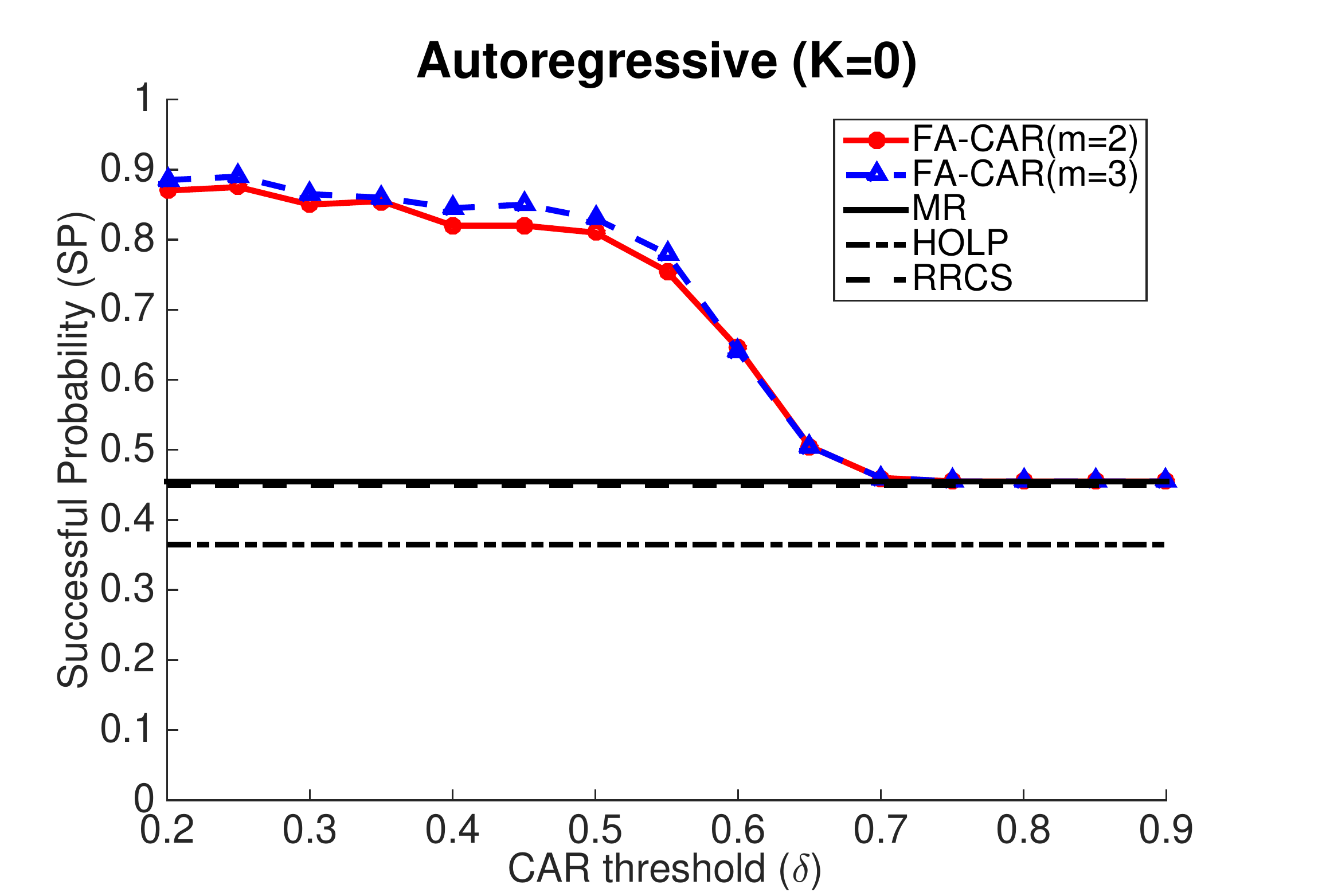} 
\includegraphics[width=0.45\textwidth,height=0.35\textwidth]{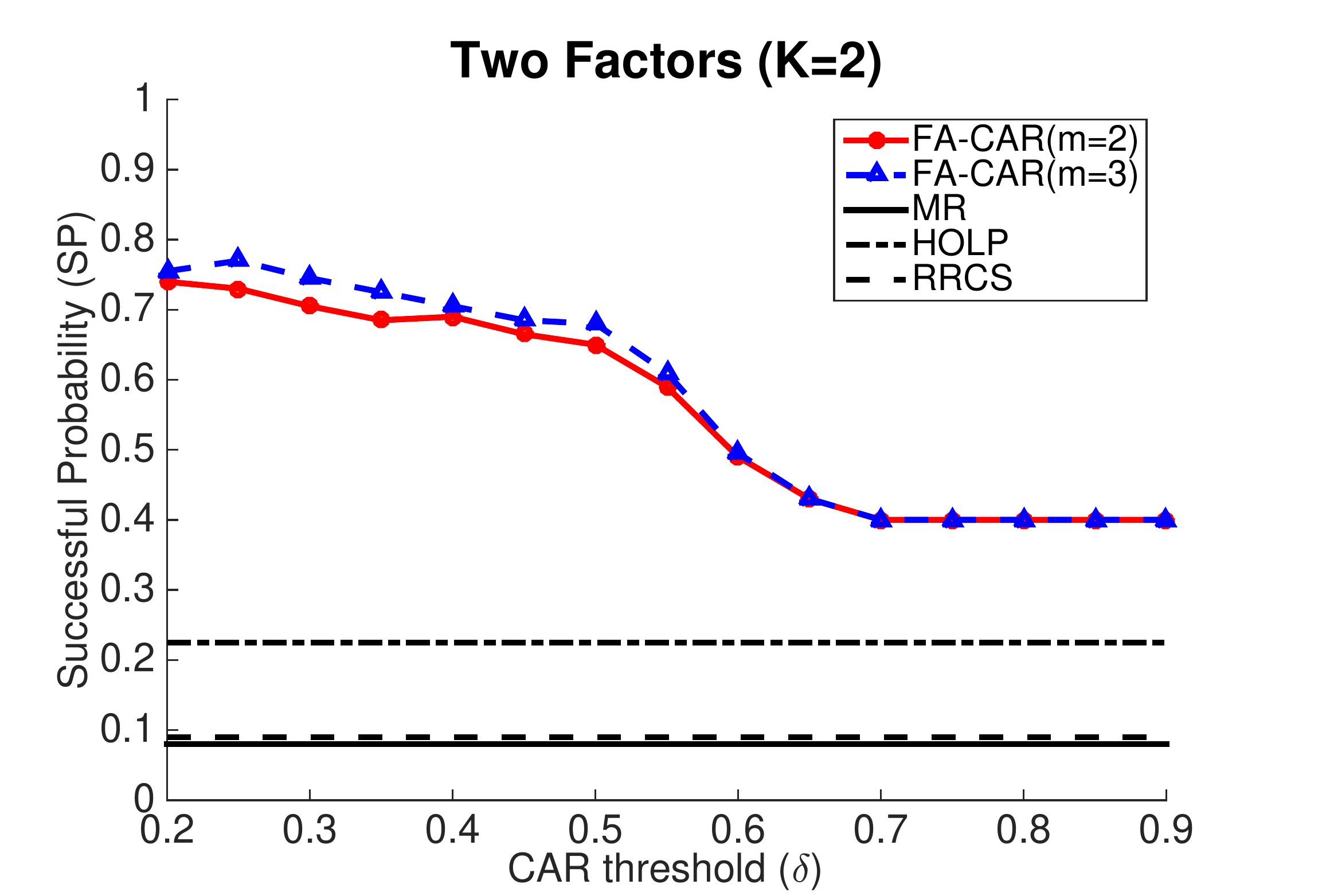} 
\caption{Experiment 3: sensitivity to tuning parameters. The ideal choice of $K$ is $K=0$ for the autoregressive design and $K=2$ for the two-factor design.}\label{fig:robust1}
\end{figure}
\begin{figure}
\includegraphics[width=0.45\textwidth,height=0.35\textwidth]{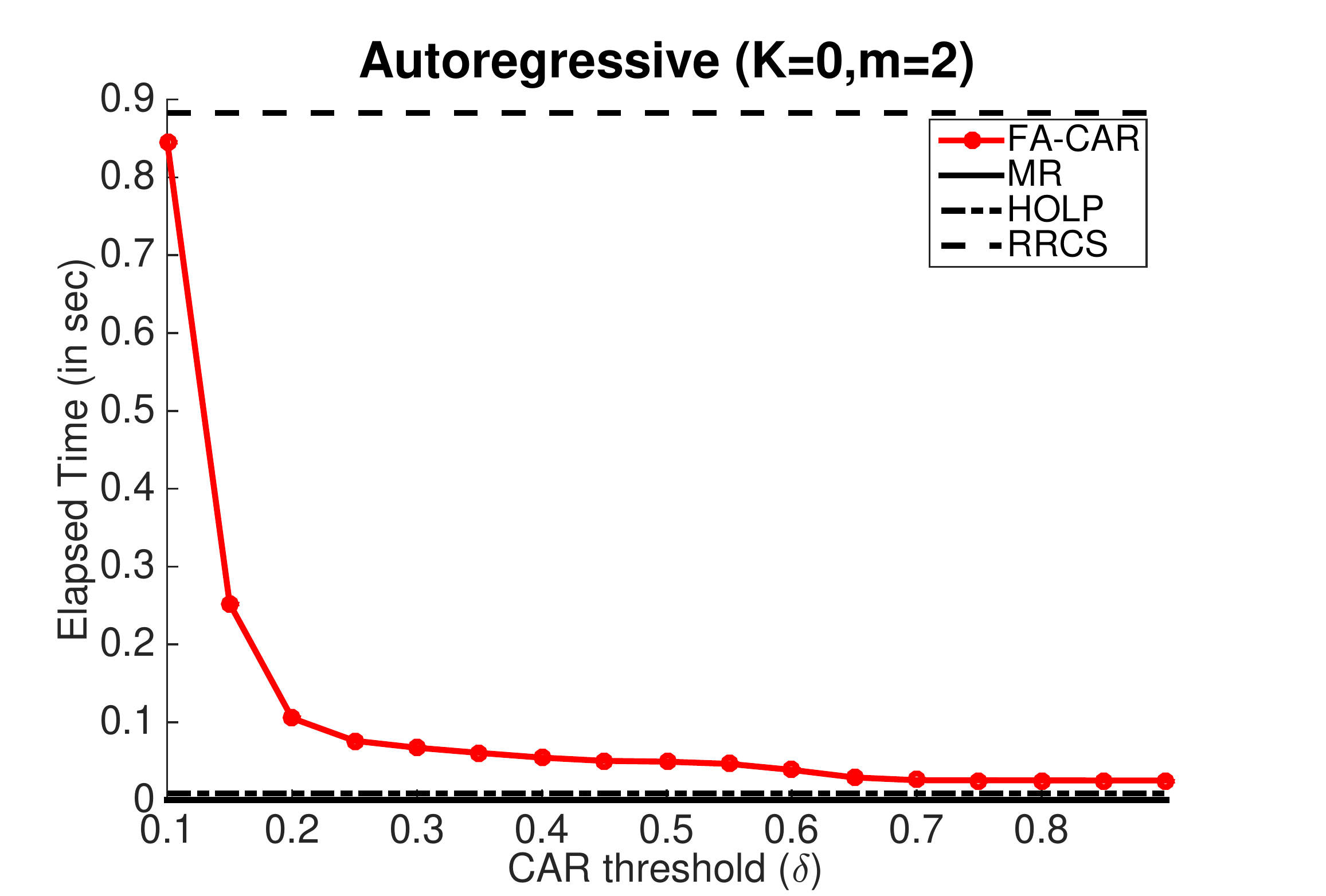} 
\includegraphics[width=0.45\textwidth,height=0.35\textwidth]{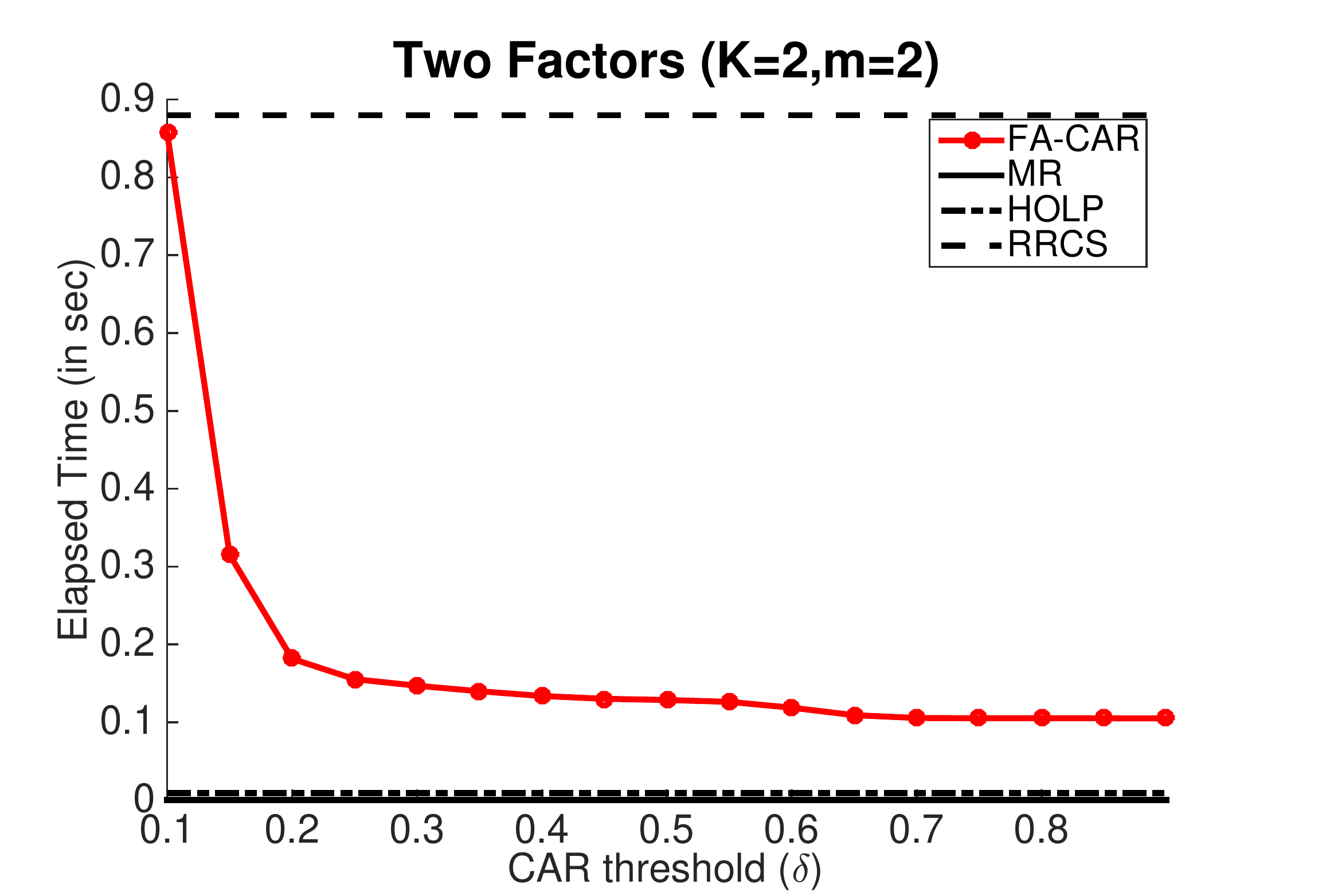} \\
\caption{Computing time in Experiment 3.}\label{fig:robust2}
\end{figure}

\paragraph{Experiment 3: Sensitivity to tuning parameters.}
We study how the performance of FA-CAR changes as the tuning
parameters $(K, \delta,m)$ vary. We fix
$(n,p,\eta,s)=(200,1000,0.5,5)$, and focus on the autoregressive
designs and two-factor designs. We implement FA-CAR for $K\in
\{0,1,2,3\}$, $m\in \{2,3\}$ and $\delta\in \{.2, .25, .3, .35,
\cdots, .9\}$. The results are shown in Figures~\ref{fig:robust1}; to save space, we only report the sure screening probability (SP). We also report the computing time for different values of $\delta$ in Figure~\ref{fig:robust2}.

{\it Choice of $K$}. 
The top two panels of Figure~\ref{fig:robust1} suggest that overshooting of
$K$ makes almost no difference in the performance, but
undershooting of $K$ could render the performance worse (e.g.,
$K=1$ for the two factors design). Even with an
undershooting $K$, FA-CAR still significantly outperforms MR and
RRCS, and is comparable with HOLP for a wide range of $\delta$.

{\it Choice of $m$}. The bottom two panels of
Figure~\ref{fig:robust1} suggest that increasing $m$ from $2$ to $3$ slightly improves
the performance especially when $\delta$ is small, but we pay a
price in computational cost. In general, $m=2$ is a good choice. 

{\it Choice of $\delta$}. From Figure~\ref{fig:robust2}, smaller $\delta$ tends to yield better
performance of FA-CAR; but as long as $\delta<0.5$, the performance
is more or less similar (and is much better than the other 
methods). From 
Figure~\ref{fig:robust2}, the computing time decreases as
$\delta$ increases. Combining Figures~\ref{fig:robust1}-\ref{fig:robust2}, we find that
$\delta=0.5$ achieves a good balance between statistical
accuracy and computational cost.

\section{Extension to generalized linear models}  \label{sec:GLM} 
In bioinformatics and machine learning, it is often the case that the responses are not continuous, and the generalized linear models (GLM) is more appropriate for modeling the data. Consider a GLM with the canonical link: The responses  $y_1,\ldots,y_n$ are independent of each other, and each $y_i$ has a probability density from the exponential family: 
\[
f(y_i) = \exp\{y_i\theta_i - b(\theta_i) + c(y_i)\},  \qquad 1\leq i\leq n, 
\]
where $b(\cdot)$ and $c(\cdot)$ are known functions satisfying that $E[y_i]=b'(\theta_i)$. The parameter $\theta_i$ is called the canonical or natural parameter. GLM models that
\[
\theta_i = \theta_i(X_i) = \beta_0 + X_i'\beta, \qquad 1\leq i\leq n. 
\]
The parameters $\beta_0\in\mathbb{R}$ and $\beta\in\mathbb{R}^p$ are unknown.\footnote{We can also add a dispersion parameter $\sigma^2$ and all the results continue to hold.} 
Same as before, we call a nonzero entry of $\beta$ a ``signal". We are interested in ranking the variables such that the top ranked variables contain as many signals as possible. 

Marginal Ranking (MR) can be conveniently extended to GLM, where the marginal correlation coefficients  are replaced by the maximum marginal likelihoods or maximum marginal likelihood estimators  \citep{fan2009ultrahigh,fan2010sure}. 
With similar ideas, FA-CAR can also be extended to GLM. 

Write $X=[X_1,\ldots,X_n]'=[x_1,\ldots,x_p]$. Our main ideas are as follows:
\begin{itemize}
\item In the FA step, we run SVD on $X$ and obtain $\tilde{X}=[\tilde{x}_1,\ldots,\tilde{x}_p]$ same as before. We then treat the left singular vectors $\hat{u}_1,\ldots,\hat{u}_K$ as ``confounding" variables and create a new GLM, where the response $y$ is the same but the variables are $\hat{u}_1,\ldots,\hat{u}_K, \tilde{x}_1,\ldots,\tilde{x}_p$. 
\item In the CAR step, we construct the graph $\cG^\delta$ same as before. We then modify the scores $T_{j|\call}$ using some local log-likelihood ratios and combine these modified $T_{j|\call}$'s to get $T_j^*$ similarly as before. 
\end{itemize}

We now describe the GLM version of FA-CAR. In the FA step, let $X=
\sum_{k=1}^n \hat{\sigma}_k \hat{u}_k \hat{v}_k'$ be the SVD of $X$ and define 
\beq
\tilde{X}= X -\sum_{k=1}^K \hat{\sigma}_k\hat{u}_k\hat{v}_k'. 
\eeq
Write $\hat{U}=[\hat{u}_1,\cdots,\hat{u}_K]$. Let $\hat{U}'_i$ and $\tilde{X}'_i$ be the $i$-th row of $\hat{U}$ and $\tilde{X}$, respectively, $1\leq i\leq n$. We consider a new GLM where $y_1,\ldots,y_n$ are independent and each $y_i$ has the probability density
\beq \label{GLM-postFA}
f(y_i) = \exp\{y_i\tilde{\theta}_i - b(\tilde{\theta}_i) + c(y_i)\},  \qquad \tilde{\theta}_i = \beta_0+ \hat{U}_i'\alpha + \tilde{X}_{i}'\beta. 
\eeq
The log-likelihood of the new GLM is 
\[
\ell(\beta_0, \beta; y, \tilde{X}, \hat{U}) = \sum_{i=1}^n \big[ \big(\beta_0+\hat{U}_i'\alpha+\tilde{X}_i'\beta\big)y_i - b\big(\beta_0+\hat{U}_i'\alpha+\tilde{X}_i'\beta\big) + c(y_i)\big]. 
\]
In the special case of linear models (we assume $\beta_0=0$), Model \eqref{GLM-postFA} becomes $y=\hat{U}\alpha+\tilde{X}\beta + {\cal N}(0,I_n)$. Since $\alpha\in\mathbb{R}^K$ is low-dimensional and the columns of $\hat{U}$ are orthogonal to the columns of $\tilde{X}$, we can regress $y$ on $\hat{U}$ only to get the least-squares estimator $\hat{\alpha}^{ols}$ and subtract $\hat{U}\hat{\alpha}^{ols}$ from $y$. This gives $\tilde{y}$. So we have recovered Model \eqref{linearmod3}.

In the CAR step, we introduce a ``local log-likelihood" for each subset $V\subset\{1,\ldots,p\}$:   
\[
\ell(\alpha, \beta_{0}, \beta_V; y,\tilde{X}, \hat{U})= \sum_{i=1}^n \big[ \big(\beta_0+ \hat{U}_i'\alpha + \tilde{X}_{i,V}'\beta_V\big)y_i  - b\big(\beta_0+ \hat{U}_i'\alpha + \tilde{X}_{i,V}'\beta_V\big) + c(y_i)\big], 
\]
where $\tilde{X}_{i,V}$ is obtained from restricting $\tilde{X}_i$ to the coordinates in $V$. 
Define the maximum partial log-likelihood as
\[ 
\hat{\ell}_V(y, \tilde{X}, \hat{U}) =\max_{\alpha, \beta_0, \beta_V} \ell(\alpha, \beta_{0}, \beta_V; y,\tilde{X}, \hat{U}). 
\]
This quantity $\hat{\ell}_V(y, \tilde{X}, \hat{U})$ serves as a counterpart of $\|P_V\tilde{y}\|^2$ in the case of linear models.  
We then introduce a counterpart of $T_{j|\call}$ for GLM:
\beq  \label{Tjcall-GLM}
T^{glm}_{j|\call } = \hat{\ell}_{\call}(y;
\tilde{X}, \hat{U}) - \hat{\ell}_{\call\setminus\{j\}}(y, \tilde{X}, \hat{U}). 
\eeq
Let $\cG^\delta$ and $A_{\delta,j}(m)$ be the same as in \eqref{def-graph} and \eqref{Def-Aj}. The final scores are 
\beq \label{GLM-Score}
T_j^* = \max \big\{T^{glm}_{j|\call}: \call\in {\cal A}_{\delta,j}(m)\big\}.
\eeq

We use a numerical example to compare FA-CAR with two GLM versions of MR: MR-1  
\citep{fan2009ultrahigh} uses the 
maximum marginal likelihood estimator to rank variables, 
and 
MR-2 \citep{fan2010sure} uses the maximum marginal log-likelihood to rank variables. We are not aware of any direct extensions of HOLP and RRCS for GLM, so we omit the comparison with them. Fixing
$(n,p,\eta,s)=(200,1000,3,5)$, we generate the designs similarly as in Section~\ref{sec:simu} and generate binary $y_i$'s using the logistic regression setting. In Table~\ref{logistictable}, we report the performance of three methods, where the measures are the same as those in Experiment 2 in Section~\ref{sec:simu}. It suggests a significant advantage of FA-CAR over the other two methods.

\begin{table}[h]
\center \caption{Comparison of ranking methods for the logistic regression. The measures, SP, Type II, and Size, are defined the same as those in Table~\ref{lineartable}.}
\label{logistictable} \scalebox{.9}{
\begin{tabular}{c|c|c c c}
  \hline
  \multirow{2}{*}{Designs} & \multirow{2}{*}{Measure} & \multicolumn{3}{|c|}{Method} \\
  \cline{3-5}
   & & FA-CAR & MR-1 & MR-2 \\
 \hline
\multirow{2}{*}{Tridiagonal} & SP/Type II & \textbf{0.87/0.15} & 0.49/0.62 & 0.49/0.62 \\
  \cline{3-5}
  & Size & \textbf{13} & 225 & 226.5  \\
 \hline
\multirow{2}{*}{Autoregressive} & SP/Type II & \textbf{0.84/0.20} & 0.55/0.59 & 0.55/0.59 \\
  \cline{3-5}
  & Size & \textbf{11.5} & 160 & 159.5  \\
 \hline
  \multirow{2}{*}{Equal corr.} & SP/Type II & \textbf{0.26/1.10} & 0.05/2.03 & 0.05/2.02 \\
  \cline{3-5}
  & Size & \textbf{510} & 977 & 980 \\
 \hline
\multirow{2}{*}{Two factors} & SP/Type II & \textbf{0.86/0.22} & 0.13/1.95 & 0.13/1.94 \\
  \cline{3-5}
  & Size & \textbf{22} & 709.5 & 707.5  \\
 \hline

\end{tabular}}
\end{table}

\section{Discussions} \label{sec:discuss}
We propose a two-step method FA-CAR for variable ranking. The FA step uses PCA to create a new linear model with a sparse Gram matrix, and the CAR step exploits local covariate structures for variable ranking. Compared with the popular Marginal Ranking methods which completely ignore covariate structures, FA-CAR shows an advantage in both theory and numerical performance. At the same time, FA-CAR keeps some nice properties of MR, such as being easy to use, computationally fast, and easily extendable.  

The variable ranking is often a critical ``first step" for statistical analysis. Once we have good rankings scores for variables, we can proceed to other tasks such as deciding a proper cut-off, performing careful variable selection, and conducting follow-up lab experiments. In genetic and genomic applications, the first step of ranking often has a huge impact on the final outcomes \citep{chen2007gene}. Therefore, we believe that FA-CAR has a great potential for real applications. 

Our work is motivated by improving Marginal Ranking and is connected to the literatures on marginal screening (e.g., \cite{fan2016interaction, li2012feature,song2014varying,xue2011sure}). The ranking problem is especially interesting and challenging when the signals and noise are merely inseparable, so our work is on a high level connected to the literatures on global testing (e.g., 
\cite{ji2014rate,chen2010two}).  

FA-CAR can be used as a variable selection method if combined with a proper threshold on the scores and a good post-screening cleaning methods. If we combine FA-CAR with the covariate-assisted cleaning method in \cite{KJF-CASE}, we conjecture that it achieves the optimal rate of convergence of the Hamming selection errors for the approximate factor designs considered in this paper. The study of Hamming errors requires substantial efforts, and we leave it for future work. 

It is an interesting yet open question how to control the family-wise error rate and false discovery rate (FDR) based on FA-CAR scores. One possible approach is to compute the $p$-values associated with these scores using the theoretical or bootstrap null distributions and plug them into existing FDR controlling procedures. Another possibility is to borrow the recent ideas of controlling FDR directly in the variable selection procedure \citep{barber2015controlling}. We leave such investigations for future work. 
 
We have introduced an extension of FA-CAR to generalized linear models. This GLM version has nice numerical performance.  It is of great interest to study its theoretical properties in the future work.

\section{Proofs}  \label{sec:proofs}

\subsection{Proof of Lemma~\ref{lem:block-rates}}
We shall prove the following lemma, and Lemma~\ref{lem:block-rates} follows immediately. 
\begin{lemma} \label{lem:block-rates2}
Suppose the conditions of Lemma~\ref{lem:block-rates} hold. 
For all methods, 
\[
\eta^*(\vartheta,r,h)= 1-\min\{\vartheta,\; q^*(\vartheta,r,h)\},
\]
where  (notation: $a^2_+=\max\{a,0\}^2$ for any $a\in\mathbb{R}$)
\begin{align*}
q^{*}_{LSR}(\vartheta,r,h) &= (\sqrt{(1-h^2)r}-\sqrt{1-\vartheta})_+^2,\cr
q^{*}_{MR}(\vartheta,r,h) &= 
\begin{cases}
(\sqrt{r}-\sqrt{1-\vartheta})_+^2, & \vartheta\geq 1/2,\\
\min\bigl\{ (\sqrt{r}-\sqrt{1-\vartheta})_+^2,\;  \bigl((1-|h|)\sqrt{r}-\sqrt{1-2\vartheta}\bigr)_+^2\bigr\}, & \vartheta<1/2,
\end{cases}\cr
q^{*}_{CAR}(\vartheta,r,h) &=  
\begin{cases}
(\sqrt{r}-\sqrt{1-\vartheta})_+^2, & \vartheta\geq 1/2,\\
\min\bigl\{ (\sqrt{r}-\sqrt{1-\vartheta})_+^2,\;  \bigl(\sqrt{(1-h^2)r}-\sqrt{1-2\vartheta}\bigr)_+^2\bigr\}, & \vartheta<1/2.
\end{cases}
\end{align*}
\end{lemma}

We now prove Lemma~\ref{lem:block-rates2}. 
Consider using $t_p(q)=2q\sigma^2\log(p)$ to threshold 
\beq \label{lem-block2-scores}
\mbox{$n^{-1}|(x_j,y)|^2$ in MR},\quad \mbox{$T_j^*$ in CAR},\quad \mbox{and $n(1-h^2)|\hat{\beta}^{ols}_j|^2$ in LSR}. 
\eeq
We claim that, for all three methods,  
\beq \label{lem-block2-key}
FP_p(t_p(q)) = L_p p^{1-\rho_1(q; \vartheta,r,h)}, \qquad 
FN_p(t_p(q)) = L_p p^{1-\rho_2(q; \vartheta,r,h)},
\eeq
where the exponents are 
\begin{align*} 
& \rho_1^{MR}(q)=\rho_1^{CAR}(q)= \min\{q,\; \vartheta + (\sqrt{q}-|h|\sqrt{r})_+^2\},\cr
 & \rho_1^{LSR}(q)= q,\cr 
& \rho_2^{MR}(q)= \min\{\vartheta + (\sqrt{r}-\sqrt{q})_+^2, \; 2\vartheta + [(1-|h|)\sqrt{r}-\sqrt{q}]_+^2\},\cr
& \rho_2^{CAR}(q)= \min\{\vartheta + (\sqrt{r}-\sqrt{q})_+^2, \; 2\vartheta + [\sqrt{(1-h^2)r}-\sqrt{q}]_+^2\}, \cr
& \rho_2^{LSR}(q)= \vartheta + (\sqrt{(1-h^2)r}-\sqrt{q})_+^2. 
\end{align*} 
Given \eqref{lem-block2-key}, for each of the three methods, the quantity $q^*=q^*(\vartheta,r,h)$ in Lemma~\ref{lem:block-rates2} is the solution of $\rho_2(q)=1$. As a result, $FN_p(t_p(q))\to 0$ for any $q<q^*$, and $FN_p(t_p(q))\to \infty$ for any $q>q^*$. It follows that
\[
SS^*_p= s_p + FP_p(t_p(q^*)) = p^{1-\vartheta} + L_pp^{1-\rho_1(q^*)}= L_pp^{1-\min\{\vartheta,q^*\}}. 
\] 
Here the last equality comes from the expressions of $\rho_1(q)$ for all three methods. This proves Lemma~\ref{lem:block-rates2}.

It remains to prove \eqref{lem-block2-key}. Let $M_j$ be a symbol that represents the scores in \eqref{lem-block2-scores} for each method. For an even $j$, define the events:
\begin{align*}
B_{j1}= \{\beta_{j-1}=0,\beta_j\neq 0, M_j<t_p(q)\},\;\; B_{j2} = \{\beta_{j-1}\neq 0,\beta_j\neq 0, M_j<t_p(q)\},\cr
D_{j1}= \{\beta_{j-1}=0,\beta_j=0, M_j>t_p(q)\},\;\; D_{j2} = \{\beta_{j-1}\neq 0,\beta_j=0, M_j>t_p(q)\}.
\end{align*} 
There is a false negative at location $j$ over the events $B_{j1}$ and $B_{j2}$, and there is a false positive over the events $D_{j1}$ and $D_{j2}$. We can similarly define these four events for an odd $j$, by replacing $(j-1)$ by $(j+1)$. It is seen that
\[
FN_p(t_p(q))=\sum_{j=1}^p \bigl[\mathbb{P}(B_{j1})+\mathbb{P}(B_{j2})\bigr], \qquad
FP_p(t_p(q))=\sum_{j=1}^p \bigl[\mathbb{P}(D_{j1})+\mathbb{P}(D_{j2})\bigr],
\]
Therefore, to show \eqref{lem-block2-key}, it suffices to calculate the probabilities of the above events. We only consider an even $j$, and the case for an odd $j$ is similar. 

First, consider MR, where the score for variable $j$ is $M_j=n^{-1}|(x_j,y)|^2$. Note that
\[
n^{-1/2}(x_j'y) = \cN\left(\sqrt{n}(h\beta_{j-1}+\beta_j),\; \sigma^2\right). 
\]
So $M_j/\sigma^2$ has a non-central chi-square distribution with the non-centrality parameter 
equal to $n\sigma^{-2}|h\beta_{j-1}+\beta_j|^2$. On the event $B_{j1}$, $n\sigma^{-2}|h\beta_{j-1}+\beta_j|^2=n\sigma^{-2}\beta_j^2 = 2r\log(p)$. It follows that 
\begin{align*}
\mathbb{P}(B_{j1}) &= \eps_p(1-\eps_p) \cdot \mathbb{P}\left( \chi^2_1(2r\log(p))< 2q\log(p)\right)\cr
&= \eps_p(1-\eps_p)  \cdot L_pp^{-(\sqrt{r}-\sqrt{q})_+^2} = L_p p^{-\vartheta -(\sqrt{r}-\sqrt{q})_+^2}.
\end{align*}
Here, the second equality is due to Mills' ratio and elementary properties of non-central chi-square distributions. 
On the event $B_{j2}$, if $h\geq 0$, 
\[
n\sigma^{-2}|h\beta_{j-1}+\beta_j|^2 = \begin{cases}
(1+|h|)^2\cdot 2r\log(p), & \text{if } \mathrm{sign}(\beta_{j-1})=\mathrm{sign}(\beta_{j}),\\
(1-|h|)^2\cdot 2r\log(p), & \text{if } \mathrm{sign}(\beta_{j-1})\neq \mathrm{sign}(\beta_{j}).\\
\end{cases}
\]
If $h<0$, we have similar results except that the two cases swap. As a result,
\begin{align*}
\mathbb{P}(B_{j2}) &= (\eps^2_p/2) \cdot \mathbb{P}\left( \chi_1^2\bigl(2r(1+|h|)^2\log(p)\bigr)< 2q\log(p)\right)\cr
&+(\eps^2_p/2) \cdot \mathbb{P}\left( \chi_1^2\bigl(2r(1-|h|)^2\log(p)\bigr)< 2q\log(p)\right)\cr
&=L_p p^{-2\vartheta - [(1+|h|)\sqrt{r}-\sqrt{q}]_+^2} + L_p p^{-2\vartheta - [(1-|h|)\sqrt{r}-\sqrt{q}]_+^2}\cr
&= L_p p^{-2\vartheta - [(1-|h|)\sqrt{r}-\sqrt{q}]_+^2}. 
\end{align*}
Combining the above results, we have found that
\[
FN_p(t_p(q))  = \sum_{j=1}^p L_p p^{- \min\left\{  \vartheta + (\sqrt{r}-\sqrt{q})_+^2,\; 2\vartheta + [(1-|h|)\sqrt{r}-\sqrt{q}]_+^2 \right\}} = L_pp^{1-\rho_2^{MR}(q)}. 
\]
Similarly, on the event $D_{j1}$, $n\sigma^{-2}|h\beta_{j-1}+\beta_j|^2=0$, and on the event $D_{j2}$, $n\sigma^{-2}|h\beta_{j-1}+\beta_j|^2=h^2\cdot 2r\log(p)$. We then have 
\begin{align*}
\mathbb{P}(D_{j1}) & = (1- \eps_p)^2 \cdot \mathbb{P}\left( \chi_1^2(0)> 2q\log(p)\right)=L_pp^{-q},\cr
\mathbb{P}(D_{j2}) &= \eps_p(1- \eps_p) \cdot \mathbb{P}\left( \chi_1^2(2rh^2\log(p))> 2q\log(p)\right)= L_p p^{-\vartheta -(\sqrt{q}-|h|\sqrt{r})_+^2}.
\end{align*}
As a result, 
\[
FP_p(t_p(q))  = \sum_{j=1}^p L_p p^{- \min\left\{q,\;   \vartheta + (\sqrt{q}-|h|\sqrt{r})_+^2 \right\}} = L_pp^{1-\rho_1^{MR}(q)}. 
\]

Next, consider LSR. The score $M_j=n(1-h^2)|\hat{\beta}^{ols}_j|^2$. The least squares estimator satisfies that $\hat{\beta}=(X'X)^{-1}X'y \sim \cN(\beta, n^{-1}\sigma^2\Theta^{-1})$. Here, $\Theta$ is blockwise diagonal with two-by-two blocks. We immediately have  
\[
\hat{\beta}_{j}^{ols}\sim \cN\left(\beta_j,\; \frac{\sigma^2}{n(1-h^2)}\right). 
\]
So $M_j/\sigma^2$ has a non-central chi-square distribution with the non-centrality parameter $n(1-h^2)\sigma^{-2}\beta_j^2$. On both of the events $B_{j1}$ and $B_{j2}$, the non-centrality parameter is equal to $(1-h^2)\cdot 2r\log(p)$; it is easy to see that the probability of $B_{j1}$ dominates. 
It follows that
\begin{align*}
FN_p(t_p(q)) &= L_p \sum_{j=1}^p 
\mathbb{P}(B_{j1}) = L_p p\eps_p\cdot \mathbb{P}\left( \chi_1^2(2r(1-h^2)\log(p))< 2q\log(p)\right)\cr
&=L_pp^{1-\vartheta - (\sqrt{(1-h^2)r}-\sqrt{q})_+^2}= L_pp^{1-\rho_2^{LSR}(q)}. 
\end{align*}
On both of the events $D_{j1}$ and $D_{j2}$, the non-centrality parameter is equal to $0$, and the probability of $D_{j1}$ dominates. It follows that
\begin{align*}
FP_p(t_p(q)) &= L_p \sum_{j=1}^p 
\mathbb{P}(D_{j1}) = L_p p\cdot \mathbb{P}\left( \chi_1^2(0)> 2q\log(p)\right)\cr
&=L_pp^{1- q}= L_pp^{1-\rho_1^{LSR}(q)}. 
\end{align*}

Last, consider CAR. Note that $T_j^*=\max\{T_{j|\{j\}}, T_{j|\{j-1,j\}}\}$. It is easy to see that $T_{j|\{j\}}$ coincides with the score in MR.
To obtain the distribution of $T_{j|\{j-1,j\}}$, we apply \eqref{Tjcall-illust}. Let $\eta=(x_{j-1}'y, x_j'y)'$ and $H$ be the two-by-two matrix with unit diagonals and off-diagonals $h$. It follows from \eqref{Tjcall-illust} that
\[
T_{j|\{j-1,j\}} = n^{-1}(\eta'H^{-1}\eta-\eta_1^2) = \frac{1}{n(1-h^2)}(\eta_2-h\eta_1)^2. 
\]
Write $W=\frac{1}{\sqrt{n(1-h^2)}}(\eta_2-h\eta_1)$. Then, $T_{j|\{j-1,j\}}=W^2$. 
Since $\eta\sim \cN(nH\beta, nH)$, 
\[
W\sim \cN\left( \sqrt{n(1-h^2)}\beta_j, \; \sigma^2\right). 
\]
To summarize, we have found that 
\begin{align} \label{lem-block2-CARdist}
& T_{j|\{j\}}/\sigma^2\sim \chi_1^2\bigl( n\sigma^{-2}|\beta_{j}+h\beta_{j-1}|^2  \bigl),\cr
&   
T_{j|\{j-1,j\}}/\sigma^2\sim \chi_1^2\bigl( n\sigma^{-2}(1-h^2)\beta_j^2\bigr).
\end{align}
Consider the type II errors. We use a simply fact that $\max\{T_{j|\{j\}}, T_{j|\{j-1,j\}}\}<t_p(q))$ has a probability that is upper bounded by either the probability of $T_{j|\{j\}}<t_p(q)$ or the probability of $T_{j|\{j-1,j\}}<t_p(q)$, so we can take the minimum of these two probabilities as an upper bound. On the event $B_{j1}$, the non-centrality parameters for the two statistics are $n\beta_j^2$ and $n(1-h^2)\beta_j^2$. Therefore, the type II error is determined by the behavior of $T_{j|\{j\}}$. It follows that
\begin{align*}
\mathbb{P}(B_{j1})& \leq \eps_p(1-\eps_p)\cdot \mathbb{P}(T_{j|\{j\}}<t_p(q))\cr
& = \eps_p(1-\eps_p) \cdot \mathbb{P}\bigl(\chi^2_1( 2r\log(p) )<t_p(q) \bigr)\cr
& = L_pp^{-\vartheta-(\sqrt{r}-\sqrt{q})_+^2}. 
\end{align*}
On the event $B_{j2}$, the non-centrality parameter for $T_{j|\{j-1,j\}}/\sigma^2$ is the same as before, which is $n(1-h^2)\sigma^{-2}\beta_j^2=(1-h^2)\cdot 2r\log(p)$. The non-centrality parameter for $T_{j|\{j\}}/\sigma^2$ has been studied in the MR case, which is equal to $(1\pm |h|)^2\cdot 2r\log(p)$. In the case of $(1+|h|)^2\cdot 2r\log(p)$, since $(1+|h|)^2\geq 1-h^2$, the type II error is determined by the behavior of $T_{j|\{j\}}$. In the case of  $(1-|h|)^2\cdot 2r\log(p)$, since $1-h^2\geq (1-|h|)^2$, the type II error is determined by the behavior of $T_{j|\{j-1,j\}}$. As a result,
\begin{align*}
\mathbb{P}(B_{j2})&\leq (\eps_p^2/2) \cdot \mathbb{P}\bigl(T_{j|\{j\}}<t_p(q)\bigr) +  (\eps_p^2/2)  \cdot \mathbb{P}\bigl(T_{j|\{j-1,j\}}<t_p(q)\bigr)\cr
& = (\eps_p^2/2)  \cdot  \mathbb{P}\bigl(\chi^2_1( 2r(1+|h|)^2\log(p) )<t_p(q) \bigr)\cr
& \;\; + (\eps_p^2/2)  \cdot  \mathbb{P}\bigl(\chi^2_1( 2r(1-h^2)\log(p) )<t_p(q) \bigr)\cr
& = L_p p^{-2\vartheta - [ (1+|h|)\sqrt{r}-\sqrt{q}]_+^2} + L_p p^{-2\vartheta - (\sqrt{(1-h^2)r}-\sqrt{q})_+^2}\cr
& = L_p p^{-2\vartheta - (\sqrt{(1-h^2)r}-\sqrt{q})_+^2}. 
\end{align*}
Combining the above results, we have 
\[
FN_p(t_p(q))  = \sum_{j=1}^p L_p p^{- \min\left\{  \vartheta + (\sqrt{r}-\sqrt{q})_+^2,\; 2\vartheta + (\sqrt{(1-h^2)r}-\sqrt{q})^2 \right\}} = L_pp^{1-\rho_2^{CAR}(q)}. 
\]
Consider the type I errors. On the event $D_{j1}$, both non-centrality parameters in \eqref{lem-block2-CARdist} become $0$. We then use the probability union bound to get
\begin{align*}
\mathbb{P}(D_{j1}) & \leq (1-\eps_p)^2\cdot \bigl[\mathbb{P}( T_{j|\{j\}}>t_p(q) ) + \mathbb{P}(T_{j|\{j-1,j\}}>t_p(q)) \bigr]\cr
&= (1-\eps_p)^2 \cdot 2\mathbb{P}\bigl(\chi^2_1( 0 )>t_p(q) \bigr)\cr
&=L_pp^{-q}. 
\end{align*}
Similarly, on the event $D_{j2}$, 
\begin{align*}
\mathbb{P}(D_{j2}) & \leq \eps_p (1-\eps_p)\cdot \bigl[\mathbb{P}( T_{j|\{j\}}>t_p(q) ) + \mathbb{P}(T_{j|\{j-1,j\}}>t_p(q)) \bigr]\cr
&=  \eps_p (1-\eps_p) \cdot \left[ \mathbb{P}\bigl(\chi^2_1( 2h^2r\log(p) )>t_p(q) \bigr) + \mathbb{P}\bigl(\chi^2_1( 0 )>t_p(q) \bigr) \right]\cr
&=L_pp^{-\vartheta-(\sqrt{q}-|h|\sqrt{r})_+^2} + L_pp^{-\vartheta - q}. 
\end{align*}
It follows that 
\[
FP_p(t_p(q))  = \sum_{j=1}^p L_p p^{- \min\left\{ q,\;  \vartheta + (\sqrt{q}-|h|\sqrt{r})_+^2 \right\}} = L_pp^{1-\rho_1^{CAR}(q)}. 
\]
The proof is now complete.

\subsection{Proof of Corollaries~\ref{cor:no-cancellation}-\ref{cor:compare}}
Consider Corollary~\ref{cor:no-cancellation}. It suffices to prove
\beq \label{cor1-key}
\omega_j(r,m)\geq c_0r, \qquad \mbox{for all }j\in S. 
\eeq 
Once \eqref{cor1-key} is true, the $q^*(\vartheta,r,m)$ defined in Theorem~\ref{thm:SS} satisfies $q^*(\vartheta,r,m)\geq (\sqrt{c_0r}-\sqrt{1-\vartheta})_+^2$. Then, Corollary~\ref{cor:no-cancellation} follows. 

We show \eqref{cor1-key}. 
Fix $j\in S$ and let $\call_k$ be the unique component of $\cG^\delta_{S}$ that contains $j$. By \eqref{cond-graphlet} and \eqref{choose-delta+m}, $\call_k\in\cA_{\delta,j}(m)$. It then follows from \eqref{omega_j(r,m)} that $\omega_{j}(r,m)\geq \omega_{j|\call_k}(r)$. Furthermore, by arguments in Lemma~\ref{A.2}, $\omega_{j|\call_k}(r)=\omega^*_j(r) + o(\omega^*_j(r))$. Combining the above gives 
\[
\omega_{j}(r,m)\gtrsim \omega^*_{j}(r) = \frac{n A^0_{j|\call_k}}{2\sigma^2\log(p)}\beta_j^2. 
\]
Note that $A_{j|\call_k}^0=G_0^{j,j}-G_0^{j,N}(G_0^{N,N})^{-1}G_0^{N,j}$, where $N=\call_k\setminus\{j\}$. We arrange indices in $\call_k$ such that $j$ is the first index. By the matrix inverse formula, $A^0_{j|\call_k}$ is the inverse of the first diagonal of $(G_0^{\call_k,\call_k})^{-1}$. As a result, 
\[
A^0_{j|\call_k}\geq \lambda_{\min}(G_0^{\call_k,\call_k})\geq c_0,
\]
where the last inequality comes from $G_0\in {\cal M}_p(g, \gamma, c_0, C_0)$ and $|\call_k|\leq \ell_0\leq g$. Combining it with $|\beta_j|\geq \tau_p$ gives \eqref{cor1-key}. 

Consider Corollary~\ref{cor:compare}. In FA-MR, since the columns of $\tilde{X}$ have unequal norms, we first normalize them: $\tilde{x}_j^*=(\sqrt{n}/\|\tilde{x}_j\|) \tilde{x}_j$. We then rank variables by the marginal correlation coefficients
\[
|(\tilde{x}_j^*, \tilde{y})|/(\tilde{x}_j^*, \tilde{x}_j^*) = n^{1/2}|(\tilde{x}_j/\|\tilde{x}_j\|, \tilde{y})| = n^{1/2} \|P_{\{j\}}\tilde{y}\|, 
\]
where we recall that $P_{\{j\}}\tilde{y}$ is the projection of $\tilde{y}$ onto $\tilde{x}_j$. 
So FA-MR is a special case of FA-CAR with $m=1$. The claim then follows from the fact that $\omega_{j}(r,m)$ is a monotone increasing function of $m$.

\subsection{Proof of Lemma~\ref{lem:Eigvec1}}
Without loss of generality, we assume $v_1'\hat{v}_1\geq 0$. 
By definition, $\Theta \hat{v}_1=\hat{\lambda}_1\hat{v}_1$, where
$\Theta=\lambda_1v_1v_1' + G_0$. It follows that \beq
\label{lem-Eig1-0} \lambda_1(v_1'\hat{v}_1) v_1 + G_0\hat{v}_1 =
\hat{\lambda}_1 \hat{v}_1. \eeq By Weyl's inequality,
$|\hat{\lambda}_1-\lambda_1| \leq \|G_0\| \leq \|G_0\|_{\infty}
\leq \lambda_1/3$. As a result, \beq \label{lem-Eig1-1} (2/3)
\lambda_1\leq \hat{\lambda}_1\leq (4/3) \lambda_1. \eeq In
particular, the minimum eigenvalue of $\hat{\lambda}_1 I_p - G_0$
is lower bounded by $(2/3) \lambda_1 - \|G_0\|\geq \lambda_1/3$.
So $(\hat{\lambda}_1 I_p - G_0)$ is always positive definite. So
we can solve from \eqref{lem-Eig1-0} to get \beq
\label{lem-Eig1-2} \hat{v}_1=(I_p - \hat{\lambda}_1^{-1}G_0)^{-1}
\cdot \frac{\lambda_1 (v_1'\hat{v}_1)}{\hat{\lambda}_1} v_1. \eeq

We now show the claim.
Write $\Delta = (I_p - \hat{\lambda}_1^{-1}G_0)^{-1}-I_p$ and $\epsilon=\frac{\lambda_1 (v_1'\hat{v}_1)}{\hat{\lambda}_1}-1$.
We have
\begin{align}\label{lem-Eig1-3}
\|\hat{v}_1&-v_1\|_\infty = \big\| (I_p +\Delta) (1+\epsilon) v_1- v_1 \big\|_\infty\cr
& \leq \|\Delta v_1\|_\infty + \|\epsilon v_1 + \epsilon \Delta v_1\|_\infty \cr
&\leq  \|\Delta\|_\infty \|v_1\|_\infty +|\epsilon| \cdot \bigl(\|v_1\|_\infty + \|\Delta \|_\infty \|v_1\|_\infty \bigr).
\end{align}

First, we bound $\|\Delta\|_\infty$. Since $(\Delta +I_p) (I_p - \hat{\lambda}_1^{-1}G_0)=I_p$, we have
\[
\Delta= \hat{\lambda}_1^{-1}G_0 +  \Delta \hat{\lambda}_1^{-1}G_0.
\]
Using the triangular inequality, $\|\Delta\|_\infty \leq \hat{\lambda}_1^{-1}\|G_0\|_\infty + \|\Delta\|_\infty \hat{\lambda}_1^{-1}\|G_0 \|_\infty$. It follows that
\[
\|\Delta\|_\infty \leq  \frac{\hat{\lambda}_1^{-1} \|G_0\|_\infty}{1-\hat{\lambda}_1^{-1}
\|G_0\|_\infty}.
\]
By assumption, $\|G_0\|_\infty\leq \lambda_1/3$; by \eqref{lem-Eig1-1}, $\hat{\lambda}^{-1}\leq \frac{3}{2}\lambda^{-1}_1$. So the denominator $1-\hat{\lambda}_1^{-1}
\|G_0\|_\infty\geq 1/2$. It follows that
\beq  \label{lem-Eig1-5}
\|\Delta\|_\infty\leq 3\lambda_1^{-1}\|G_0\|_\infty.
\eeq

Next, we bound $|\epsilon|$. Note that
\[
|\epsilon|= |\frac{\lambda_1 (v_1'\hat{v}_1)}{\hat{\lambda}_1}-1|
\leq |1-\frac{\lambda_1}{\hat{\lambda}_1}| + \frac{\lambda_1}{\hat{\lambda}_1}\cdot |1-v_1'\hat{v}_1|
\leq \frac{3}{2} \frac{\|G_0\|}{\lambda_1} +  \frac{3}{2} |1-v_1'\hat{v}_1|,
\]
where we use $|\hat{\lambda}_1-\lambda_1|\leq \|G_0\|$ and \eqref{lem-Eig1-1} in the last inequality. We now consider $|1-v_1'\hat{v}_1|$. Multiplying both sides of
\eqref{lem-Eig1-0} by $\hat{v}_1'$ from the left, we get $\lambda_1(v_1'\hat{v}_1)^2 + \hat{v}_1'G_0\hat{v}_1 = \hat{\lambda}_1$. So
\[
(v_1'\hat{v}_1)^2 = \frac{\hat{\lambda}_1}{\lambda_1} - \frac{\hat{v}_1'G_0\hat{v}_1}{\lambda_1}.
\]
As a result,
\[
|1-v_1'\hat{v}_1| \leq 1- (v_1'\hat{v}_1)^2 \leq |1-\frac{\hat{\lambda}_1}{\lambda_1}|  + \frac{ |\hat{v}_1'G_0\hat{v}_1|}{\lambda_1}\leq  \frac{\|G_0\|}{\lambda_1}  + \frac{\|G_0\|}{\lambda_1} = 2\frac{\|G_0\|}{\lambda_1}.
\]
Combining the above gives
\beq \label{lem-Eig1-4}
|\epsilon|\leq \frac{9}{2} \lambda_1^{-1}\|G_0\|\leq \frac{9}{2} \lambda_1^{-1}\|G_0\|_\infty.
\eeq

We plug \eqref{lem-Eig1-5}-\eqref{lem-Eig1-4} into
\eqref{lem-Eig1-3}, and use $\|G_0\|_\infty\leq \lambda_1/3$. It yields
\[
\|\hat{v}_1-v_1\|_\infty \leq \|v_1\|_\infty \left( \frac{3\|G_0\|_\infty}{\lambda_1} +  \frac{9\|G_0\|_\infty}{2\lambda_1}(1+\frac{3\|G_0\|_\infty}{\lambda_1})   \right)\leq\frac{12  \|v_1\|_\infty\|G_0\|_\infty}{\lambda_1}.
\]
This proves the claim.
%
%

\subsection{Proof of Theorem~\ref{thm:Eigen}}
As preparation, we introduce $\tv_k$, defined in \eqref{thm-Eigen-2}, as a counterpart of $\hat{v}_k$ for $1\leq k\leq K$.
By Weyl's inequality, for any $1\leq k\leq K$, $|\hat{\lambda}_k - \lambda_k| \leq \|G_0\| \leq \|G_0\|_{\infty} \leq C_1^{-1}\lambda_K\leq C_1^{-1}\lambda_k$. It follows that
\beq \label{thm-Eigen-1}
 \frac{C_1-1}{C_1}
\lambda_k \leq \hat{\lambda}_k \leq \frac{C_1+1}{C_1} \lambda_k
\eeq
Write $\Lambda=\mathrm{diag}(\lambda_1,\cdots,\lambda_K)$ and $\hat{\Lambda}=\mathrm{diag}(\hat{\lambda}_1,\cdots,\hat{\lambda}_K)$. Recall that $V=[v_1,\cdots,v_K]$ and $\hat{V}=[\hat{v}_1,\cdots,\hat{v}_K]$. 
By definition,
\[
\hat{\lambda}_k \hat{v}_k = \Theta \hat{v}_k = (V\Lambda V' + G_0) \hat{v}_k ,
\]
which implies $(\hat{\lambda}_k I_p -G_0)\hat{v}_k  = V \Lambda V' \hat{v}_k$. By \eqref{thm-Eigen-1}, $(\hat{\lambda}_k I_p -G_0)$ is positive definite. Hence,
\beq \label{thm-Eigen-2}
\hat{v}_k = (I_p - \hat{\lambda}_k^{-1} G_0 )^{-1}\tilde{v}_k, \qquad \mbox{where}\quad \tilde{v}_k\equiv \hat{\lambda}_k^{-1} (V\Lambda V')\hat{v}_k.
\eeq
Write $\tilde{V}=[\tilde{v}_1,\cdots,\tilde{v}_K]$.

We now show the first claim about $\|\hat{V}\hat{V}'-VV'\|_{\max}$. It is seen that
\begin{align} \label{thm-Eigen-3}
\|\hat{V}\hat{V}'-& VV'\|_{\max} \leq \|\hat{V}\hat{V}'-\tilde{V}\tilde{V}'\|_{\max} + \|\tilde{V}\tilde{V}' - VV'\|_{\max}\cr
&\leq  \sum_{k=1}^K \|\hat{v}_k\hat{v}_k' - \tilde{v}_k\tilde{v}_k' \|_{\max} + \|\tilde{V}\tilde{V}' - VV'\|_{\max}
\equiv I + II.
\end{align}
First, we bound $I$. For $1\leq k\leq K$, letting $\Delta_k=(I_p -\hat{\lambda}^{-1}_k G_0 )^{-1} - I_p$, we have $\|\hat{v}_k-\tv_k\|_{\infty}\leq \|\Delta_k\|_\infty \|\tilde{v}_k\|_\infty$ and
\begin{align} \label{thm-Eigen-4}
 \|\hat{v}_k\hat{v}_k' - \tilde{v}_k\tilde{v}_k' \|_{\max} & \leq
\|\hat{v}_k-\tv_k\|_{\infty}^2
+ 2\|\hat{v}_k-\tv_k\|_{\infty}\|\tv_k\|_{\infty}\cr
& \leq \|\tv_k\|^2_\infty (\|\Delta_k\|_\infty^2 + 2\|\Delta_k\|_\infty)
\end{align}
We consider $\|\Delta_k\|_\infty$ and $\|\tilde{v}_k\|_\infty$ separately.
Observing that $\Delta_k= \hat{\lambda}_k^{-1}G_0 +  \Delta_k \hat{\lambda}_1^{-1}G_0 $, we apply the triangle inequality to get $\|\Delta_k\|_\infty\leq \hat{\lambda}_k^{-1}\|G_0\|_\infty + \|\Delta_k\|_\infty  \hat{\lambda}_k^{-1}\|G_0 \|_\infty$. By \eqref{thm-Eigen-1} and the assumption, $\hat{\lambda}_k^{-1}\|G_0 \|_\infty\leq \frac{C_1}{C_1-1}\lambda_k^{-1}\|G_0\|_\infty\leq \frac{1}{C_1-1}$. It follows that
\[
\|\Delta\|_\infty \leq \frac{\hat{\lambda}_k^{-1}\|G_0\|_\infty}{1-\hat{\lambda}_k^{-1}\|G_0\|_\infty}\leq \frac{C_1-1}{C_1-2}\frac{\|G_0\|_\infty}{\hat{\lambda}_k}\leq \frac{C_1}{C_1-2}\frac{\|G_0\|_\infty}{\lambda_k}.
\]
Recalling that $\tilde{v}_k=\hat{\lambda}_k^{-1}V\Lambda V'\hat{v}_k$, we have $\|\tilde{v}_k\|_\infty \leq \hat{\lambda}_k^{-1} \|V\|_{\max} \|\Lambda V'
\hat{v}_k\|_1 \leq \hat{\lambda}_k^{-1}\|V\|_{\max}\cdot \sqrt{K} \|\Lambda V'\hat{v}_k\|$. Since $\|V\|=1$ and $\|\hat{v}_k\|=1$, $\|\Lambda V'\hat{v}_k\|\leq \lambda_1$. It follows that
\[
\|\tilde{v}_k\|_\infty\leq
\sqrt{K}\Big(\frac{\lambda_1}{\hat{\lambda}_k}\Big) \|V\|_{\max}\leq \sqrt{K}\frac{C_1}{C_1-1}\Big(\frac{\lambda_1}{\lambda_k}\Big) \|V\|_{\max}.
\]
Combining the above with \eqref{thm-Eigen-4} and noting that $\|\Delta_k\|^2_\infty\leq C\|\Delta_k\|_\infty$, we find that
\beq \label{thm-Eigen-5}
 \|\hat{v}_k\hat{v}_k' - \tilde{v}_k\tilde{v}_k' \|_{\max} \leq C\lambda_k^{-1} \Big(\frac{\lambda_1}{\lambda_k}\Big)^2\|V\|^2_{\max}\|G_0\|_\infty.
\eeq

We then bound $II$. It is seen that
\begin{align} \label{thm-Eigen-6}
\tilde{V}\tilde{V}' - VV' & = \sum_{k=1}^K \hat{\lambda}_k^{-2} V\Lambda V'\hat{v}_k\hat{v}_k' V\Lambda V' - VV'\cr
&= V\Lambda V' \hat{V}\hat{\Lambda}^{-2}\hat{V}'V\Lambda V'  - VV'\cr
&= V(M'M-I_K)V', \qquad \mbox{where}\quad M \equiv \hat{\Lambda}^{-1}\hat{V}'V\Lambda.
\end{align}
We now derive a bound for $\|M'M-I_K\|$. By definition,
\[
(V\Lambda V'+G_0)\hat{V} = \Theta\hat{V} = \hat{V}\hat{\Lambda}.
\]
Multiplying both sides by $V'$ from the left and noting that $V'V=\hat{V}\hat{V}'=I_K$, we find that $\Lambda (V'\hat{V})+V'G_0\hat{V} = (V'\hat{V})\hat{\Lambda}$. This yields an equation for $\hat{V}'V$:
\[
(\hat{V}'V)\Lambda= \hat{\Lambda}(\hat{V}'V) - \hat{V}'G_0V.
\]
As a result, we can write
\beq \label{thm-Eigen-7}
M = \hat{\Lambda}^{-1}\big[\hat{\Lambda}(\hat{V}'V) - \hat{V}'G_0V\big] = \hat{V}'V - \hat{\Lambda}^{-1}(\hat{V}'G_0V).
\eeq
Write $B=-\hat{\Lambda}^{-1}(\hat{V}'G_0V)$. It follows from \eqref{thm-Eigen-7} that
\begin{align*}
\|M'M-I_K\| &= \|(\hat{V}'V+B)'(\hat{V}'V+B) -I_K \|\cr
& \leq \|V' \hat{V}\hat{V}'V - I_K \| + 2\|B\|\|V'\hat{V}\|+\|B\|^2\cr
&\leq \| V'(\hat{V}\hat{V}'-VV')V\| + (2\|B\|+\|B\|^2)\cr
&\leq \| \hat{V}\hat{V}'- VV' \| + (2\|B\|+\|B\|^2),
\end{align*}
where the third inequality is because $\|V'\hat{V}\|\leq 1$ and
$V'V=I_K$. Applying the sine-theta theorem
\citep{davis1970rotation}, we obtain $\|\hat{V}\hat{V}'- VV' \|
\leq \frac{\|G_0\|}{\lambda_K - \|G_0\|}$. Combining it with
$\|G_0\|\leq C_1^{-1}\lambda_K$ gives $\|\hat{V}\hat{V}'- VV' \|
\leq \frac{C_1}{C_1-1}\lambda_K^{-1}\|G_0\|$. Moreover, $\|B\|\leq
\hat{\lambda}_K^{-1}\|G_0\| \leq
\frac{C_1}{C_1-1}\lambda_K^{-1}\|G_0\|$ by \eqref{thm-Eigen-1}. We
plug these results into the above inequality and find that \beq
\label{thm-Eigen-8} \|M'M-I_K\|\leq C\lambda_K^{-1}\|G_0\|. \eeq
Combining \eqref{thm-Eigen-8} with \eqref{thm-Eigen-6} gives 
\begin{align} \label{thm-Eigen-9} 
\|\tilde{V}\tilde{V}' - VV'\|_{\max} & \leq
\|V(M'M-I_K)\|_{\infty}\|V'\|_{\max}\cr
& \leq K\sqrt{K}\|M'M-I_K\|
\|V\|_{\max}^2 \cr
&\leq C\lambda_K^{-1}\|G_0\|\|V\|_{\max}^2. 
\end{align}

We plug \eqref{thm-Eigen-5} and \eqref{thm-Eigen-9} into \eqref{thm-Eigen-3}, and note that $\|G_0\|\leq \|G_0\|_\infty$ and $\lambda_k\geq \lambda_K$ for all $1\leq k\leq K$. It follows that
\[
\|\hat{V}\hat{V}' - VV'\|_{\max} \leq C_2\lambda_K^{-1} \Big(\frac{\lambda_1}{\lambda_K}\Big)^2\|V\|^2_{\max}\|G_0\|_\infty
\]
This proves the first claim.

We then show the second claim about $\|G-G_0\|_{\max}$. Note that
\begin{align} \label{thm-Eigen-10}
\|G - G_0\|_{\max} & = \|\hat{V}\hat{\Lambda}\hat{V}' - V\Lambda V'\|_{\max}\cr
& \leq \|\hat{V}\hat{\Lambda}\hat{V}' - \tilde{V}\hat{\Lambda}\tilde{V}'\|_{\max} + \| \tilde{V}\hat{\Lambda}\tilde{V}' - V\Lambda V'\|_{\max}\cr
&\leq \sum_{k=1}^K \hat{\lambda}_k \|\hat{v}_k\hat{v}_k' - \tv_k\tv_k'\|_{\max} + \| \tilde{V}\hat{\Lambda}\tilde{V}' - V\Lambda V'\|_{\max}\cr
&\leq C\sum_{k=1}^K \Big(\frac{\lambda_1}{\lambda_k}\Big)\|V\|^2_{\max}\|G_0\|_\infty + \| \tilde{V}\hat{\Lambda}\tilde{V}' - V\Lambda V'\|_{\max},
\end{align}
where we have used \eqref{thm-Eigen-5} and \eqref{thm-Eigen-1} in the last inequality. It remains to bound $\| \tilde{V}\hat{\Lambda}\tilde{V}' - V\Lambda V'\|_{\max}$. We recall the definition of $\tv_k$ in \eqref{thm-Eigen-2} and $M$ in \eqref{thm-Eigen-6}.
By direct calculations,
\begin{align*}
\|\tilde{V}\hat{\Lambda}\tilde{V}' - V\Lambda V'\|_{\max} &=  \big\| \sum_{k=1}^K \hat{\lambda}^{-1}_k (V\Lambda V')\hat{v}_k \hat{v}_k' (V\Lambda V') - V\Lambda V'\big\|_{\max}\cr
&= \| V\Lambda V'(\hat{V}\hat{\Lambda}^{-1}\hat{V}' )V\Lambda V' - V\Lambda V'\|_{\max}\cr
&= \| V\Lambda (V' \hat{V}M)V'  - V\Lambda V'\|_{\max} \cr
&\leq K \sqrt{K} \|V\|_{\max}^2 \|\Lambda\| \| V' \hat{V}M- I_K\|.
%
\end{align*}
By \eqref{thm-Eigen-7}, $M=\hat{V}'V+B$, where $B=-\hat{\Lambda}^{-1}(\hat{V}'G_0V)$. In the proof of \eqref{thm-Eigen-8}, we have seen that $\|\hat{V}\hat{V}'- VV' \| \leq \frac{C_1}{C_1-1}\lambda_K^{-1}\|G_0\|$ and $\|B\|\leq\frac{C_1}{C_1-1}\lambda_K^{-1}\|G_0\|$. It follows that
\begin{align*}
\| V' \hat{V}M-& I_K\|  =\|(V'\hat{V}\hat{V}'V - I_K)  + V'\hat{V}B  \|\cr
&\leq \|\hat{V}\hat{V}' - VV'\| + \|B\| \leq C\lambda_K^{-1}\|G_0\|.
\end{align*}
Combining the above gives
\beq \label{thm-Eigen-11}
\| \tilde{V}\hat{\Lambda}\tilde{V}' - V\Lambda V'\|_{\max} \leq C \Big(\frac{\lambda_1}{\lambda_k}\Big) \|V\|_{\max}^2\|G_0\|
\eeq
We plug \eqref{thm-Eigen-11} into \eqref{thm-Eigen-10}, and note that $\|G_0\|\leq \|G_0\|_\infty$ and $\lambda_k\geq \lambda_K$ for all $1\leq k\leq K$. It yields that
\[
\|G - G_0\|_{\max}\leq C_2' \Big(\frac{\lambda_1}{\lambda_k}\Big)^2 \|V\|_{\max}^2\|G_0\|_\infty.
\]
This proves the second claim.

\subsection{Proof of Lemma~\ref{lem:FA}}
By \eqref{cond-design} and Weyl's inequality, we know
\[
\hat{\sigma}_K^2/n \geq \lambda_K - \|G_0\| \gg \log(p) \quad \hat{\sigma}_{K+1}^2/n \leq
\|G_0\| \ll \log(p)
\]
and hence $\hat{K}_p = K$ where $\hat{K}_p$ is defined in
\eqref{choose-K}.

For any $1\leq i\leq K$ and $1\leq j\leq p$, since $\Theta(j,j) =
G_0(j,j) + \sum_{k=1}^K \lambda_k v_k(j)^2 = 1$, we have
\[
\lambda_K  v_i(j)^2 \leq \lambda_i v_i(j)^2 \leq \sum_{k=1}^K \lambda_k
v_k(j)^2 =1 - G_0(j,j) \leq 1
\]
and hence by \eqref{cond-design}, we have
\[
\max_{1\leq k\leq K}\|v_k\|_{\infty}^2 \leq \lambda_K^{-1} =
o(1/\max\{s_p,\log(p)\})
\]
Then we can get the desired result by directly applying Theorem~\ref{thm:Eigen}.

\subsection{Proof of Lemma~\ref{lem:statistics}} \label{subsec:preliminary}
We give several technical lemmas that are used frequently in the main proofs. Lemma~\ref{A.1} gives the exact distribution of the statistic $T_{j|\call}$. Lemma~\ref{A.2} implies that, if we replace $\cG^{\delta}_{0,S}$ with $\cG^\delta_S$ in the definition of $\omega_{j|\call}(r)$, the resulting change is negligible. Lemma~\ref{A.3} shows that those small entries of $G$ (recall that $G$ is the Gram matrix of model \eqref{linearmod3}) has negligible effects on screening. In this section, we write $G(j,j)=G^{j,j}$ and $G^{\{j\},N}=G^{j,N}$ for notation convenience, similarly for $G_0^{j,j}$ and $G_0^{j,N}$.

\begin{lemma}\label{A.1}
Under the conditions of Theorem~\ref{thm:SS}, for
$\call\subset \{1,\cdots,p\}$ such that $|\call |\leq g$ and any $j\in \call$, $T_{j|\call}$ has the same distribution as $W^2$,  where $W\sim \mathcal{N}(w,\sigma^2)$, 
\[
w=n^{1/2}
A_{j|\call}^{1/2}\left[ \beta_j + A_{j|\call}^{-1}
(G^{j,\call^c}\beta^{\call^c} - 
G^{j,N}(G^{N,N})^{-1}G^{N,\call^c}\beta^{\call^c})\right], 
\]
and $A_{j|\call}=G^{j,j}-G^{j,N}(G^{N,N})^{-1}G^{N,j}$ with
$N=\call\setminus \{j\}$.
\end{lemma}


If there is an edge between $i$ and $j$ in $\cG^\delta$, then $|G_0(i,j)|\geq |G(i,j)|-\|G-G_0\|_{\max}\geq \delta - o(\delta_p)\gtrsim 1.01\delta_p$, where we have used Lemma~\ref{lem:FA} and the assumption \eqref{choose-delta+m}. So there must be an edge between $i$ and $j$ in $\cG_0^\delta$. In other words, $\cG_S^\delta$ is a subgraph of $\cG_{0,S}^\delta$ by removing some edges. Fix $j\in S$ and $\call \subset \call^{(j)}$ where $\call^{(j)}$ is the unique component of $\cG_{0,S}^{\delta}$  that contains $j$.  We introduce a counterpart of $\omega_{j|\call}(r)$ in \eqref{omega_j_I} when $\call \subset \cG_{S}^{\delta}$:
\[
\tilde{\omega}_{j|\call}(r) =  \frac{n A_{j|\call}}{2\sigma^2\log(p)} \left\{  \beta_j + A_{j|\call}^{-1}
[G^{j,F} - 
G^{j,N}(G^{N,N})^{-1}G^{N,F}]\beta^{F} \right\}^2. 
\]
where $A_{j|\call}=G^{j,j}-G^{j,N}(G^{N,N})^{-1}G^{N,j}$ with
$N=\call\setminus \{j\}$ and $F=\call^{(j)}\setminus\call$.

\begin{lemma}\label{A.2}
Under conditions of Theorem~\ref{thm:SS}. For any $j\in S$, if  $\call^{(j)}$, the unique component of $\cG^\delta_{0,S}$ that contains $j$, has a size $\leq g$, then for any $\call \subset \call^{(j)}$ we have $A_{j|\call}^0 \geq c_0$ and  $|A_{j|\call}^0 - A_{j|\call \cap \cG^\delta_{S}}| = o(\delta_p)$. Moreover, if $\call \subset \cG_{S}^{\delta}$ then $|\tilde{\omega}_{j|\call}(r)-\omega_{j|\call}(r)|/\omega_{j|\call}(r)=o(1)$.
\end{lemma}

\begin{lemma}\label{A.3}
Define the matrix $G^\delta\in\mathbb{R}^{p,p}$ by $G^{\delta}(i,j) = G(i,j)
\mathbf{1}\{|G(i,j)|>\delta\}$ for $1\leq i,j\leq p$. Under conditions of Theorem~\ref{thm:SS},  for any $\call\subset \{1,2,\dots,p\}$ and $\cJ \subset \{1,2,\dots,p\}$,
\[
\big\|\left(G^{\call,\cJ} - (G^{\delta})^{\call,\cJ}\right)\beta^{\cJ}\big\|_{\infty} \leq
C\left(\log(p)^{-(1-\gamma)} + s_p\|G-G_0\|_{\max}\right)\tau_p
 = o(\tau_p).
 \]
\end{lemma}


Now we prove Lemma~\ref{lem:statistics}. We denote $\call^{(j)}$ as $\call_k$ for some $1\leq k\leq M$. We know by Lemma~\ref{A.1} that $T_{j|\call} \sim \cN^2(w_1,\sigma^2)$
where
\[
w_1 = n^{1/2}A_{j|\call}^{1/2}\left(\beta_j + A_{j|\call_k}^{-1}[G^{j,F} - G^{j,N}(G^{N,N})^{-1}G^{N,F}]\beta^F \right) + \rom{1} + \rom{2}
\]
and
\[
\rom{1} = n^{1/2}A_{j|\call}^{-1/2}
G^{j,\call_k^c}\beta^{\call_k^c}, \rom{2} =
-n^{1/2}A_{j|\call_k}^{-1/2}
G^{j,N}(G^{N,N})^{-1}G^{N,\call_k^c}\beta^{\call_k^c}
\]

It's easy to see that there exists a constant $C$ such that
$\|G^{j,N}(G^{N,N})^{-1}\|_{\infty} \leq C$. By definition of $\call_k$ and the fact that $\cG_S^{\delta}\subset \cG_{0,S}^{\delta}$, we know
$(G^{\delta})^{\call_k,\call_k^c}\beta^{\call_k^c}=0$ where $G^\delta$ is defined in Lemma~\ref{A.3}. Hence we
have
\[
G^{\call_k,\call_k^c}\beta^{\call_k^c}  = (G\beta)^{\call_k} - G^{\call_k,\call_k}\beta^{\call_k} =
\left(G^{\call_k,\call_k^c} -
(G^{\delta})^{\call_k,\call_k^c}\right)\beta^{\call_k^c}
\]
By Lemma~\ref{A.3}, we know
\[
\|G^{\call_k,\call_k^c}\beta^{\call_k^c}\|_{\infty} = \|(G\beta)^{\call_k} - G^{\call_k,\call_k}\beta^{\call_k}\|_{\infty} = o(\tau_p) =
 o(n^{-1/2} \sqrt{\log(p)})
\]
which suggests that
\[
\max\{|\rom{1}|,|\rom{2}|\} = o(\sqrt{\log(p)})
\]
By Lemma~\ref{A.2}, we know
\[
w_1 = \sigma\sqrt{2\log(p)\tilde{\omega}_{j|\call}(r)} + o(\sqrt{\log(p)}) = \sigma\sqrt{2\log(p)\omega_{j|\call}(r)} + o(\sqrt{\log(p)})
\]
which implies Lemma~\ref{lem:statistics}.

\subsection{Proof of Theorem~\ref{thm:typeI+II}}

For any $j\in S$ and any $\call\in{\cal A}_{\delta,j}(m)$ with $\call \subset \call^{(j)}$, we know by Lemma~\ref{lem:statistics} and Mill's ratio, 
\begin{eqnarray*}
P(T_{j|\call}\leq t_p(q)) & \lesssim &
P\left(|\cN(\sqrt{2\omega_{j|\call}(r)\sigma^2\log(p)},\sigma^2)|\leq
\sqrt{2q\sigma^2\log(p)}\right) \\ & \leq & P\left(\cN(0,1)\geq
\sqrt{2\omega_{j|\call}(r)\log(p)}-\sqrt{2q\log(p)}\right)\\ & \leq & L_p
p^{-[(\sqrt{\omega_{j|\call}(r)}-\sqrt{q})_+]^2}
\end{eqnarray*}
which implies that 
\[
P(T_j^* \leq t_p(q)) \leq \min_{\call \in {\cal A}_{\delta,j}(m),\call \subset \call^{(j)}} P(T_{j|\call} \leq t_p(q)) \leq L_p p^{-[(\sqrt{\omega_j(r,m)}-\sqrt{q})_+]^2}
\]

Therefore, we get 
\[
E( |S\setminus \hat{S}(q)|)  = \sum_{j\in S}P(T_j^* \leq t_p(q)) \leq  L_p \sum_{j\in S} p^{-[(\sqrt{\omega_j(r,m)}-\sqrt{q})_+]^2}
\]

Now we look at the second term. Recall $G^{\delta}$ defined in Lemma~\ref{A.3}. We
show that each row of $G^{\delta}$ has at most
$C(\log(p))^{\gamma}$ nonzeros for some constant $C$.
For any $1\leq i\leq p$, suppose there are $K_i$ nonzeros at $i$th
row of $G^{\delta}$. By Lemma~\ref{lem:FA}, when p is sufficiently large
$\|G-G_0\|_{\max} \leq \delta/2 $. Hence we have
\[
C_0 \geq \sum_{j=1}^p |G_0(i,j)|^{\gamma} = \sum_{j=1}^p
|G(i,j)-(G(i,j)-G_0(i,j))|^{\gamma} \geq K_i
(\delta-\delta/2)^{\gamma}
\]
which implies that
\[
K_i \leq C_0 (\delta/2)^{-\gamma} \leq C(\log(p))^{\gamma}
\]

By a classical result in graph theory
\citep{frieze1999splitting}, we have for any $1\leq j\leq p$,
\beq\label{graphbound} |{\cal A}_{\delta,j}(m+1)| \leq (m+1)(e\max_i K_i)^{m}
\leq C (\log(p))^{\gamma m} \eeq
Define
\[S_\delta^\ast(m) = \{1\leq j \leq p: j \ \text{is connected to } S \ \text{through a path of length} \leq m \ \text{in} \ \cG^{\delta} \}\]

For $j\in S_\delta^\ast(m)$, we know there exists a node
$i\in S\cap\call$ for some $\call\in {\cal A}_{\delta,j}(m+1)$. This implies
that $j\in \call \in {\cal A}_{\delta,j}(m+1)$. Hence we have
\[
|S_\delta^\ast(m)| \leq \sum_{i\in S} m |{\cal A}_{\delta,j}(m)|
\leq C s_p (\log(p))^{\gamma m}
\]

For $j\not\in S_\delta^\ast(m)$ and any
${\cal A}_{\delta,j}(m)$, by Lemma~\ref{A.1} we can write
$T_{j|\call} = W^2$ where $W\sim \cN(w,\sigma^2)$ and
\[
w = n^{1/2}A_{j|\call}^{-1/2} G^{j,\call^c}\beta^{\call^c} -
n^{1/2}A_{j|\call}^{-1/2}
G^{j,N}(G^{N,N})^{-1}G^{N,\call^c}\beta^{\call^c}
\]
By definition of $S_\delta^\ast(m)$, we know
$(G^{\delta})^{\call,\call^c}\beta^{\call^c}
=(G^{\delta})^{\call,S}\beta^{S}=0$. By Lemma~\ref{A.3} we have
\[
\|G^{\call,\call^c}\beta^{\call^c}\|_{\infty}
=\|\left(G^{\call,\call^c} -
(G^{\delta})^{\call,\call^c}\right)\beta^{\call^c}\|_{\infty} =
o(\tau_p)
\]
which implies that $w=o(n^{1/2}\tau_p) = o(\sqrt{\log(p)})$. Hence
we have
\[
T_{j|\call} \sim \sigma^2 \chi^2_1\left(o(\log(p))\right)
\]
which suggests
\[
P(T_{j|\call} > t_p(q)) \leq  P\left(\cN(0,1) > \sqrt{2q\log(p)} -
o(\sqrt{\log(p)})\right) \lesssim L_p p^{-q}
\]
Hence by union bound, we have
\[
P(T_j^* > t_p(q)) \leq \sum_{\call \in {\cal A}_{\delta,j}(m)} P(T_{j|\call} > t_p(q))
\leq L_p p^{-q} |{\cal A}_{\delta,j}(m)| \leq L_p p^{-q}
\]

Therefore, we have
\begin{eqnarray*}
E( |\hat{S}(q)|) & = & \sum_{j\in S^*_\delta(m)} P(T_j^*
> t_p(q)) + \sum_{j\not\in S^*_\delta(m)} P(T_j^* > t_p(q)) \\
& \leq & | S^*_\delta(m)|+ p \cdot L_p p^{-q} \\
& \leq &  C s_p (\log(p))^{\gamma m} + L_p p^{1-q}
\end{eqnarray*}
which proves the theorem.

%

\appendix  
\setcounter{equation}{0} 

\section{Proof of secondary lemmas}

\subsection{Proof of Lemma~\ref{A.1}}
We need some preparations. First, we show that
\beq\label{A-bound}
 A_{j|\call} \geq
\nu_{\min}(G^{\call,\call}) \gtrsim c_0,
\eeq
so that $A_{j|\call}$ is always positive. A helpful result is the matrix blockwise inverse fomular
\[
\begin{bmatrix}
A & B\\
C & D
\end{bmatrix}^{-1}
= \begin{bmatrix}
A^{-1}+A^{-1}BMCA^{-1} & -A^{-1}BM\\
-MCA^{-1} & M
\end{bmatrix}=\begin{bmatrix}
A^{-1} & 0\\
0 & 0
\end{bmatrix} +
\begin{bmatrix} B\\ - I\end{bmatrix}M  \begin{bmatrix} C &- I\end{bmatrix},
\]
with $M= (D-CA^{-1}B)^{-1}$. Without loss of generality, we assume $j$ is the first index in $\call$. Applying the above formula, we see that $(A_{j|\call})^{-1}$ is the $(1,1)$-th entry of
$(G^{\call,\call})^{-1}$. It follows that $A_{j|\call} \geq
\nu_{\min}(G^{\call,\call})$. Since $|\call|\leq g$, it suffices to show that
$\nu_g^*(G)\gtrsim c_0$ where $\nu_g^*(G)$ is the same as in Section~\ref{subsec:condition}. For any $g\times g$
matrix $\tE$ which is a principal submatrix of $G$, let $E_0$ be
the corresponding principal submatrix of $G_0$. We know
$\nu_{\min}(E_0) \geq c_0$. By Weyl's inequality and Lemma~\ref{lem:FA},
\[
|\nu_{\min}(\tE) - \nu_{\min}(E_0) | \leq \|\tE-E_0\|_2 \leq g
\|\tE-E_0\|_{\max} = o(1/\log(p))
\]
which implies that $\nu_{\min}(\tE)\gtrsim c_0$ and hence
$\nu_g^*(G)\gtrsim c_0$ as $p$ goes to infinity.

Second, we introduce $y_1 = \tX'\ty$ and show that
\beq \label{y1}
y_1 \sim \cN(n G \beta, \sigma^2 n G).
\eeq
Since $\ty \sim \cN(\tX\beta, \sigma^2 H)$ where $H=I_n - \sum_{k=1}^K \hat{u}_k\hat{u}_k'$, we have
$y_1 = \tX'\ty \sim \cN(\tX'\tX\beta, \sigma^2 \tX'H\tX)$.
Noting that $\tX=HX$ and $G = (1/n) \tX'\tX$, we obtain
$\tX'H\tX = (HX)'H(HX) = X'H^2X = (HX)'(HX) = \tX'\tX = n G$. So \eqref{y1} follows.

We now show the claim. By definition,
\begin{eqnarray*}
T_{j|\call} & = & \|P_{\call}\ty\|^2 - \|P_{N}\ty\|^2 \\ & = &
\ty'\tX^{\call}\left((\tX^{\call})'\tX^{\call}\right)^{-1}(\tX^{\call})'\ty
- \ty'\tX^{N}\left((\tX^{N})'\tX^{N}\right)^{-1}(\tX^{N})'\ty \\ &
= & n^{-1} (y_1^{\call})'(G^{\call,\call})^{-1}y_1^{\call} -
n^{-1} (y_1^{N})'(G^{N,N})^{-1}y_1^{N} \\ & = &
n^{-1}(y_1^{\call})'\left( (G^{\call,\call})^{-1} - \left[
\begin{array}{cc} (G^{N,N})^{-1} & 0\\ 0& 0\end{array}\right]
\right) y_1^{\call}
\end{eqnarray*}
where we assume $j$ is the last index in $\call$ for the
presentation purpose. Applying the matrix inverse formula, we
obtain
\beq \label{Tjcall-expression}
T_{j|\call} = n^{-1}(y_1^{\call})'B'A_{j|\call}^{-1}B y_1^{\call}, \qquad B = \big[-G^{j,N}(G^{N,N})^{-1} , 1\big].
\eeq
Therefore, $T_{j|\call}=W^2$ for $W = n^{-1/2}A_{j|\call}^{-1/2} B (y_1^\call)$.

It remains to calculate the mean and variance of $W$. First, by \eqref{y1}, the variance of $W$ is $\sigma^2 A^{-1}_{j|\call}(B G^{\call,\call} B')$, where by definition of $B$ and elementary calculations, $BG^{\call,\call}B'=A_{j|\call}$. So $\mathrm{var}(W)=\sigma^2$.
Second, it is seen that $W= n^{-1/2}A_{j|\call}^{-1/2}(y_1^j -
G^{j,N}(G^{N,N})^{-1}y_1^N)$. It follows from \eqref{y1} that
\begin{align*}
E[W] &=n^{-1/2}A^{-1/2}_{j|\call}\left[ (G\beta)^j - G^{j,N}(G^{N,N})^{-1}(G\beta)^N   \right] \cr
&= n^{-1/2}A^{-1/2}_{j|\call}\left[ G^{j,\call}\beta^\call - G^{j,N}(G^{N,N})^{-1}G^{N,\call}\beta^{\call} +rem\right]\cr
&= n^{-1/2}A^{-1/2}_{j|\call}\left[G^{j,j}\beta_j + G^{j,N}\beta^N - G^{j,N}(G^{N,N})^{-1}G^{N,j}\beta_j -G^{j,N}\beta^N + rem \right]\cr
&= n^{-1/2}A^{-1/2}_{j|\call} \left[ A_{j|\call}\beta_j + rem\right],
\end{align*}
where $rem=G^{j,\call^c}\beta^{\call^c} -
G^{j,N}(G^{N,N})^{-1}G^{N,\call^c}\beta^{\call^c}$. So $E[W]=w$.

\subsection{Proof of Lemma~\ref{A.2}}
Fix $j$ and for any $\call \subset \call^{(j)}$ denote by $\tcall = \call \cap \cG^\delta_S$.  Without loss of generality, we assume $j$ is the first index of both sets $\call$ and $\tcall$. By Lemma~\ref{A.1}, we have seen that $A^{-1}_{j|\tcall}$ equals to the $(1,1)$-th entry of $(G^{\tcall,\tcall})^{-1}$; similarly, $(A_{j|\call}^0)^{-1}$ equals to the $(1,1)$-th entry of $(G_0^{\call,\call})^{-1}$. Since $|\call|\leq g$
\[
A_{j|\call}^0 \geq \lambda_{\min}( G_0^{\call,\call})\geq \nu_g^*(G_0)\geq c_0.
\]
This proves the first claim.

We now show the second claim. Since both $A_{j|\call}^0$ and $A_{j|\tcall}$ are upper bounded by some constant, it suffices to show that
\beq \label{lemA2-1}
|(A_{j|\call}^{0})^{-1} - A_{j|\tcall}^{-1}|=O(\delta_p).
\eeq
By triangular inequality,
\begin{align*}
|(A_{j|\call}^{0})^{-1} - A_{j|\tcall}^{-1}| & =|(G_0^{\call,\call})^{-1}(1,1) -
(G^{\tcall,\tcall})^{-1}(1,1)|\cr
&\leq
|(G_0^{\call,\call})^{-1}(1,1) -
(G^{\call,\call})^{-1}(1,1)| +
|(G^{\call,\call})^{-1}(1,1) -
(G^{\tcall,\tcall})^{-1}(1,1)|
\cr
&\equiv I +II.
\end{align*}

Consider $I$. First, since $|\call|\leq g$, $\|G^{\call,\call}-G_0^{\call,\call}\|\leq g\|G-G_0\|_{\max}=o(\delta_p)$ by Lemma~\ref{lem:FA}. Second, $\lambda_{\min}(G_0^{\call,\call})\geq \nu_g^*(G_0)\geq c_0$. It follows that
\begin{align} \label{lemA2-2}
I &\leq \|(G_0^{\call,\call})^{-1}-(G^{\call,\call})^{-1}\| \leq \|(G_0^{\call,\call})^{-1}\|
\| G^{\call,\call}- G_0^{\call,\call}\| \| (G^{\call,\call})^{-1} \|\cr
&\lesssim  c_0^{-2} \| G^{\call,\call}- G_0^{\call,\call}\| = o(\delta_p).
\end{align}

Consider $II$. By definition, we have $\tcall\subset\call$. If $\tcall=\call$ then $II=0$. Otherwise, write $N=\call\setminus\tcall$ and assume w.l.o.g. that the first $|\tcall|$ indices in $\call$ are from $\tcall$. Since $\tcall = \call \cap \cG^\delta_S$, there are no edges between nodes in $N$ and nodes in $\tcall$ in the graph $\cG^\delta_S$. This implies that
\[
\|G^{\tcall, N}\|_{\max}\leq \delta\leq C\delta_p.
\]
Introduce a blockwise diagonal matrix $D = \mathrm{diag}\big(G^{\tcall,\tcall}, G^{N, N}\big)$. It is seen that
\begin{align} \label{lemA2-3}
II &= |(G^{\call,\call})^{-1}(1,1) - D^{-1}(1,1)|\leq \| (G^{\call,\call})^{-1} - D^{-1}\| \cr
&\leq \| (G^{\call,\call})^{-1} \| \|D^{-1}\| \|G^{\call,\call}-D\| \lesssim c_0^{-2} \|G^{\call,\call}-D\|\cr
&= c_0^{-2} \left\| \begin{bmatrix} 0 & G^{\tcall,N}\\G^{N,\tcall} & 0\end{bmatrix}  \right\|
\leq c_0^{-2}\|G^{\tcall,N}\| \leq c_0^{-2} g\|G^{\tcall,N}\|_{\max} = O(\delta_p).
\end{align}
Combining \eqref{lemA2-2}-\eqref{lemA2-3}, we prove
\eqref{lemA2-1}.

Now suppose $\call \subset \cG^\delta_S$. We've shown that $|A_{j|\call} - A_{j|\call}^0 | = o(\delta_p)$. It suffices to show that the difference between $B^0 = G^{j,F}_0 - G^{j,N}_0(G^{N,N}_0)^{-1}G^{N,F}_0$ and $B = G^{j,F} - G^{j,N}(G^{N,N})^{-1}G^{N,F}$ is negligible.  In fact, by similar argument in \eqref{lemA2-2} we have $\|(G^{\call^{(j)},\call^{(j)}})^{-1} - (G_0^{\call^{(j)},\call^{(j)}})^{-1}\|= o(\delta_p)$.  Suppose w.l.o.g that $F\cup\{j\}$ are the first several indices of $\call^{(j)}$ where $\call^{(j)} = F\cup\{j\}\cup N$, then we know the inverse of $\tilde{B} = G^{F\cup\{j\},F\cup\{j\}} - G^{F\cup\{j\},N}(G^{N,N})^{-1}G^{N,F\cup\{j\}}$ is the upper left block of $(G^{\call^{(j)},\call^{(j)}})^{-1}$, and $B$ is a submatrix of $\tilde{B}$. We can define $\tilde{B}^0$ similarly where $B^0$ is a submatrix of $\tilde{B}^0$. By some simple algebra, we get $\|B-B^{0}\| = o(\delta_p)$ 

\subsection{Proof of Lemma~\ref{A.3}}
Recall that $S$ is the support set of $\beta$ and $|S|=s_p$. It is seen that
\begin{align*}
\big\|\big(G^{\call,\cJ} -
(G^{\delta})^{\call,\cJ}\big)\beta^{\cJ}\big\|_{\infty} & =
\big\|\big(G^{\call,\cJ\cap S} - (G^{\delta})^{\call,\cJ\cap
S}\big)\beta^{\cJ\cap S}\big\|_{\infty} \cr
 & \leq \|G^{\call, S} -
(G^{\delta})^{\call, S}\|_{\infty} \|\beta^{\cJ\cap S}\|_{\infty}\cr
&\leq a\tau_p\cdot  \|G^{\call, S} -
(G^{\delta})^{\call, S}\|_{\infty}.
\end{align*}
Therefore, to show the claim, it suffices to show that
\beq \label{lemA3-1}
\|G^{\call, S} -
(G^{\delta})^{\call, S}\|_{\infty}\leq C\left([\log(p)]^{-(1-\gamma)} + s_p\|G-G_0\|_{\max}\right).
\eeq
For any $1\leq i\leq p$, we define $I_i = \{1\leq j\leq p : |G(i,j)|\leq
\delta\}$. Then,
\begin{align*}
\|G^{\call, S} -
(G^{\delta})^{\call, S}\|_{\infty} & \leq
\max_{1\leq i\leq p}\sum_{j\in S} |G(i,j)-G^{\delta}(i,j)| = \max_{1\leq i\leq p} \sum_{j\in S\cap I_i} |G(i,j)| \cr
&\leq \max_{1\leq i\leq p} \sum_{j\in S\cap I_i} |G_0(i,j)| + \max_{1\leq i\leq p}
\sum_{j\in S\cap I_i} |G_0(i,j) - G(i,j)|\cr
&\leq  \max_{1\leq i\leq p}\sum_{j\in S\cap I_i} |G_0(i,j)| + s_p  \|G-G_0\|_{\max}.
\end{align*}
Therefore, to show \eqref{lemA3-1}, it suffices to show that for any $1\leq i\leq p$,
\beq  \label{lemA3-2}
\sum_{j\in S\cap I_i} |G_0(i,j)| \leq C[\log(p)]^{-(1-\gamma)}.
\eeq
We now show \eqref{lemA3-2}.
For any $j\in I_i$, $|G_0(i,j)|\leq |G(i,j)|+\|G-G_0\|_{\max}\leq \delta + \|G-G_0\|_{\max}$, where $\delta \leq C\delta_p=Cb/\log(p)$ by \eqref{choose-delta+m} and $\|G-G_0\|_{\max} =
o(1/\log(p))$ by Lemma~\ref{lem:FA}. Hence, $|G_0(i,j)|\leq b_1/\log(p)$ whenever $j\in I_i$, where $b_1>0$ is a constant. We have
\begin{align*}
\sum_{j\in S\cap I_i} |G_0(i,j)|& \leq  \sum_{j\in
I_i} |G_0(i,j)|^{\gamma} |G_0(i,j)|^{1-\gamma} \cr
&\leq b_1^{1-\gamma}[\log(p)]^{-(1-\gamma)}    \sum_{j\in
I_i} |G_0(i,j)|^{\gamma} \cr
&\leq b_1^{1-\gamma}[\log(p)]^{-(1-\gamma)}\cdot C_0,
\end{align*}
where we have used the assumption $G_0\in{\cal M}_p(g, \gamma,
c_0, C_0)$ in the last inequality. This proves \eqref{lemA3-2}.

\bibliography{FGS}
\bibliographystyle{imsart-nameyear}

\end{document}